\documentclass[aps,preprint,epsfig,superscriptaddress,rotate]{revtex4}
\usepackage{amsmath}
\usepackage{amsfonts}
\usepackage{amssymb}
\usepackage{graphicx}
\usepackage{bm}
\usepackage{epsfig}

\def\ie{i.e.\ }
\def\PD#1#2{\frac{\partial #1}{\partial #2}}    
\newcommand\HAH {\hat{H}}

\def\k #1{|#1\rangle}
\def\b #1{\langle #1|}

\def\mrm #1{\mathrm{#1}}

\def\mrm #1{\mathrm{#1}}

\input epsf
\begin{document}
\title{Distinguishability and Chiral Stability: \\
       Effects of Decoherence and Intermolecular Interactions}
\author{Heekyung Han}
\email[E--mail address: ]{hhan0410@gmail.com}
\affiliation{Department of Chemistry, Queen¡¯s University,
Kingston, Ontario K7L , Canada}
\author{David M. Wardlaw}
\email[E--mail address: ]{dwardlaw@uwo.ca}
\affiliation{Department of Chemistry, Queen¡¯s University,
Kingston, Ontario K7L , Canada}
\affiliation{Department of Chemistry, University of Western Ontario,
London, Ontario N6H 5B7, Canada}
\date{\today}
\begin{abstract}
We examine the effect of decoherence and intermolecular interactions
(chiral discrimination energies) on the chiral stability and the
distinguishability of initially pure versus mixed states in an open
chiral system. Under a two-level approximation for a system,
intermolecular interactions are introduced by a mean-field theory,
and interaction between a system and an environment is modeled by a
continuous measurement of a population difference between the two
chiral states. The resultant equations are explored for various
parameters, with emphasis on the combined effects of the initial
condition of the system, the  chiral discrimination energy and
the decoherence. We focus on factors affecting the distinguishability
as measured by population difference between the initially pure and
mixed states and on the chiral stability as measured by the purity
decay.
\end{abstract}
\maketitle
\section{Introduction}
\label{intro} The existence of stable chiral molecules has been
one of the most puzzling problems in quantum mechanics \cite{hund}.
Since the Hamiltonian of chiral molecules is symmetric with respect
to the parity operation, the eigenfunctions of the parity-invariant
Hamiltonian should also be parity eigenfunctions in the absence of
energy degeneracy \cite{sakurai}, i.e., the left- and right-handed
states of chiral molecules are not true stationary states. However,
some enantiomers are observed to be stable for a remarkably long
time despite their lack of parity symmetry. In the early days of
quantum-mechanics Hund formulated the problem of the stability of
chiral molecules in terms of quantum tunneling in a symmetrical
double-well potential \cite{hund}. The two chiral states are
represented by coherent superposition of the eigenstates. When
only the lowest two eigenstates, $\k{+}$ (even parity, ground)
and $\k{-}$ (odd parity) with energies $E_{+}$ and $E_{-}$, are
relevant in a description of the system, the two chiral states,
$\k{L}$ and $\k{R}$, localized around the potential minima, may
be written as
\begin{eqnarray}
\k{L} & = & \frac{1}{\sqrt{2}}(\k{+}+\k{-}),\\
\k{R} & = & \frac{1}{\sqrt{2}}(\k{+}-\k{-}).\end{eqnarray}
 Hence, if the system is initially prepared as one of the two chiral
states, say, $\k{R}$, it will oscillate between $\k{R}$ and
$\k{L}$ through the inter-well barrier. The tunneling time is
inversely proportional to the energy splitting between the two
lowest eigenstates. An exceedingly long tunneling time corresponding
to an extremely small energy splitting was thought to be responsible
for the stability of chirality. However, upon a quantitative test,
Hund's theory seems rather insufficient since the tunneling times
observed for stable chiral molecules vary in a wide range, from
sub-picosecond to the age of the universe, depending on the mass of
the particle and the nature of the potential, and yet no quantum
tunnelings are observed in these molecules.
Interest in the paradox of chiral stability was redrawn by the
discovery of weak neutral currents which violate parity symmetry at
the level of nonrelativistic quantum mechanics, and thus may lead
to stabilization of one of the two chiral configurations. However,
while theoretical calculations for this effect predict smaller
energies of predominant (L)-amino-acids and (D)-sugars than their
corresponding enantiomers, the energy difference is too small to be
detected experimentally so far \cite{weak_neutral}. On
the other hand, some have noted that a nonlinear Schr\"{o}dinger
equation does not require the eigenstates of the potential's
symmetry operators as the stationary states \cite{nonlinear1}. More
recently, Vardi showed that the effect of a difference between
homochiral interactions (the intermolecular interactions between
molecules of the same chirality) and heterochiral interactions (the
intermolecular interactions between molecules of the opposite
chirality), due to resultant nonlinearity of equations, can lead to
far more stabilized chiral states than expected for isolated molecules
\cite{vardi}. Several authors also have pointed out that the
interaction of molecules with the environment should be introduced to
explain the stability of chiral molecules. The environmental effects on
the system leads to decoherence, \ie,  coherence loss,   and thus to a suppression of coherent
tunneling oscillations. Intermolecular collisions \cite{collision},
interaction with photons \cite{photon}, and interaction with phonons
\cite{phonon} have also been employed as the origin of the dephasing
process.
Another problem related to the issue of stabilization of chiral
molecules is  controlling the chirality by exploiting lasers:
several proposals have been suggested for the preparation and
detection of the superposed chiral states utilizing phase-controlled
ultrashort laser pulses \cite{harris_cina_romero}, and growing
attention has been directed towards using lasers to manipulate the
molecular chirality \cite{control_brumer_control_ohtsuki}.
Considering that effects of the decoherence and the intermolecular
interactions on the system and thus on the control scenarios are far
from well-understood despite their expected serious impacts, it is
thus of great interest to understand the coherent tunneling dynamics
of chiral systems coupled to external environments more thoroughly.
In the related work \cite{cplhhan_wardlaw}, we considered an open
chiral system, where a two-level approximation was taken for the
system, and interaction between a system and an environment was
modeled by a continuous position measurement, an approach often used
in the studies of the quantum Zeno effect \cite{qzeno0} and in the
studies of decoherence effects in chemical reactions
\cite{hhan_jcp}. In particular, we focused on the interplay of the
initial coherences and the decoherence on the distinguishability of
initially pure versus initially mixed states as measured by a
population difference between the two chiral states, and on the
vulnerability to decoherence as measured by a purity decay. The
results provided some answers to three fundamental questions arising
in open chiral systems as follows.  1) Can we distinguish a pure
state from a mixture with same populations  as those of the pure
state? If the two chiral states are initially taken to be real for
the pure state, no distinguishability was shown to exist for all
time and any dephasing rate. Otherwise, the distinguishability exists,
although it disappears over time faster with increasing the
dephasing rate. 2) Why are some chiral systems stable for a very
long time? 3) Why are superpositions of stable chiral states not
observed experimentally? Our result suggested that the system, if
it is initially a well-localized state, would tend to preserve its
chirality when coupled to an environment that destroys left-right
coherences, provided that the interaction time scale (dephasing
time) is faster than the tunneling time scale. On the other hand,
the system, if initially a superposition of two chiral states,
would become very quickly a racemic mixture in the presence of the
very same environment.
In this paper, we extend the open chiral  system considered in Ref.
\cite{cplhhan_wardlaw} to include  chiral intermolecular
interactions which are introduced within a mean-field approximation
as used in Ref. \cite{vardi}. To better understand the combined
effects of the two interactions, namely the decoherence effect and
chiral intermolecular interactions, we first consider the system in
the presence of each interaction separately, review some results of
Ref. \cite{vardi}, \cite{cplhhan_wardlaw}, and then further obtain the
following results. 1) For a system with decoherence only and with an
initially localized state, decreasing the tunneling frequency tends
to suppress the decoherence process. It suggests that the more
classical system is less vulnerable to decoherence. 2) For a system
with chiral intermolecular interactions only, the nonzero chiral
discrimination energies can bring about the distinguishability
between the initially pure and the mixed states even when the chiral
states are taken to be real for the initially pure state, where no
distinguishability exists in the absence of chiral intermolecular
interactions. It makes it clear that Vardi's claim in the
Ref. \cite{vardi} (``{\it{$\cdots$ In the former case, spontaneous
symmetry breaking may take place, turning a nearly racemic mixture
into an optically active ensemble}}''), is unjustifiable since he
used initially pure states without any decoherence source, and thus
initially pure states should remain pure, and initially pure state
and mixed state are shown to behave differently in the presence of
intermolecular interactions. Finally, we introduce both the
decoherence and chiral intermolecular interactions into our model
system. The dephasing time is varied to be very long compared with
tunneling time. In this tunneling dominant region, the decoherence
process in the absence of chiral intermolecular interactions is
very slow, but, however, when combined with the chiral intermolecular
interactions, the decoherence process tends to be accelerated. For a
given chiral discrimination energy, we observe that, similar to the
case with decoherence only, as the dephasing rate increases, the
decoherence process becomes faster. On the other hand, for a given
dephasing rate, the decoherence process is seen to be influenced by
the initial condition of the system and the chiral discrimination
energy. That is, the vulnerability of the system to decoherence
depends on whether the initial state is strongly delocalized or
localized, and whether chiral discrimination energy favors
homochiral mixtures or heterochiral mixtures.
The organization of this work is as follows.  Section
\ref{formulation} presents formulations of a chiral system in the
presence of intermolecular interactions and decoherence effects.
Section \ref{model} gives brief descriptions of the initial
conditions, measures of the distinguishability and the coherence
loss, and computational methods. In Sec. \ref{deco_only} the
system in the presence of decoherence only is examined with a
particular focus on how the initial coherence, dephasing rate, and
tunneling frequency affect the distinguishability and the purity
decay. In Sec. \ref{inter_only} the system in the presence of
intermolecular interactions only is examined with an emphasis on
the distinguishiability  of the initially pure and mixed states,
as afforded by the intermolecular interactions. In Sec. \ref{both}
both effects, the decoherence and the intermolecular interactions
are introduced, and the combined effects are examined on the purity
decay rate and the stationary values of the population and the
coherence, with several different initial states for the system.
Finally, Sec. \ref{conclusion} provides conclusions and discuss
further studies.
\section{Formulation: chiral systems in the presence of dephasing
and intermolecular interactions}\label{formulation}
To study the behavior of chiral molecules we adopt the framework
of Hund's double-well model \cite{hund}: the chiral molecule is
characterized by a symmetric, one-dimensional double-well
potential $V(x)$ and the $x$ coordinate corresponds to the chiral
configuration. Also we take a two-level approximation for the
system, \ie, consider only the two lowest energy eigenstates, this
is valid when the energy is sufficiently low that other higher
energy eigenstates are not involved in the system dynamics. The
total wavefunction of the system may be written as
\begin{eqnarray}
\k{\Psi(t)}=a_{L}(t)\k{L}+a_{R}(t)\k{R},\label{totalw}\end{eqnarray}
The Hamiltonian for an isolated system in the two-level approximation
takes the form \begin{eqnarray}
\hat H_{0} & = & E_{+}\k{+}\b{+}+E_{-}\k{-}\b{-},\\
 & = & E_{m}(\k{R}\b{R}+\k{L}\b{L})-\delta(\k{L}\b{R}+\k{R}\b{L})\label{singleH}\end{eqnarray}
 where $E_{m}=\frac{(E_{+}+E_{-})}{2}$ and  $\delta =\frac{(E_-
-E_+)}{2}\equiv \hbar \omega$, where $\omega$ is the tunneling
frequency. Substitution of Eq.(\ref{singleH}) into the time-dependent
Schr\"{o}dinger equation $i\hbar\PD{}{t}\k{\Psi}=\hat H_{0}\k{\Psi}$
leads to a set of two linear equations for the expansion
coefficients $a_{R}$ and $a_{L}$:
\begin{eqnarray}
i\hbar\dot{a_{L}} & = & E_{m}a_{R}-\delta a_{L},\label{linearEL}\\
i\hbar\dot{a_{R}} & = & E_{m}a_{L}-\delta a_{R}.\label{linearER}
\end{eqnarray}
 The time evolution of $a_{R}$ and $a_{L}$ shows a tunneling motion
of the system between the two wells with oscillation frequency
$\omega$. Note that the tunneling time
($=2\pi/\omega=2\pi\hbar/\delta$) is inversely proportional to the
energy difference between the states $\k{+}$ and $\k{-}$.
To take into account the effect of the intermolecular interactions,
in particular, the homochiral interactions  and the heterochiral
interactions, we choose Vardi's approach \cite{vardi}. In Ref.
\cite{vardi} he employs a Hatree-Fock technique, where each molecule
sees only the mean field that is generated by the interactions of
all the other molecules. The homochiral and the heterochiral
interactions lead to two components of the self-consistent field,
$V_{hom}$ and $V_{het}$, respectively. Then, Eqs. (\ref{linearEL}),
(\ref{linearER}) become
\begin{eqnarray}
\dot{a_{L}} & = & -\frac{i}{\hbar}\left(E_{m}a_{L}-\delta
a_{R}\right)-\frac{i}{\hbar}\left(V_{hom}|a_{L}|^{2}+V_{het}|a_{R}|^{2}\right)a_{L},\label{nonlinearEL}\\
\dot{a_{R}} & = & -\frac{i}{\hbar}\left(E_{m}a_{R}-\delta
a_{L}\right)-\frac{i}{\hbar}\left(V_{hom}|a_{R}|^{2}+V_{het}|a_{L}|^{2}\right)a_{R}.\label{nonlinearER}\end{eqnarray}
 Here, the time-evolution is governed by two terms. The first comes
from the dynamics of the isolated system which generates a coherent
tunneling between two wells. The second term involving $V_{hom}$ and
$V_{het}$ is responsible for the intermolecular (hetero/homochiral)
interactions between the single molecule with the other surrounding
molecules within the mean-field approximation \cite{vardi}. Note
that introducing the intermolecular interactions results in a
nonlinear Schr\"{o}dinger equation of which the stationary states
does not need to be eigenstates of the parity operator
\cite{nonlinear1}. Vardi showed that if $V_{hom}$ and
$V_{het}$ are different, the second term involving nonlinear terms
may affect the equation of motion significantly, and can lead to
more stabilized chiral states or achiral states than expected for
the corresponding isolated molecule, depending on the initial
condition and the difference between $V_{hom}$ and $V_{het}$
\cite{vardi}. However, in the mean field approximation fluctuation
terms are ignored. These fluctuation terms arise from the
interaction between molecules and are expected to result in
decoherence. Thus, in Ref. \cite{vardi} the system does not go
through coherence loss and never becomes a mixed state when starting
from a pure state. Below we combine the intermolecular interactions
and the decoherence effect to describe the dynamics of chiral
molecules.
In order to simulate the decoherence effect, we consider the
environment surrounding the system: the environment can be thought
of as other degrees of freedom of the system or an inert gas or a
condensed phase of no chirality. We model the effects of
interactions between the system and the environment as the
continuous measurement of the system coordinate $\hat{x}$
\cite{bath_model,bath_model2}. Then the system density operator
$\hat{\rho}$, which is obtained by tracing the total density
operator over the environment, obeys the master equation
\cite{hhan_jcp},
\begin{eqnarray}
 \PD{\hat{\rho}}{t} = -\frac{i}{\hbar}[\HAH,\hat{\rho}]
- \gamma^{\prime}[\hat{x},[\hat{x},\hat{\rho}]] ,
\label{liouville}
\end{eqnarray}
with $\gamma^\prime$ being a coupling strength between the system
and the environment. The first term represents the system dynamics
in the absence of the environment, and the second term, a
multiplication of $\gamma^\prime$ and the double commutator, is
responsible for the environmental effect on the system and is
expected to destroy coherences between the eigenstates of the
position measurement operator, $\hat{x}$. This loss of coherence,
\ie, decoherence, has been argued as the essential ingredient to
reconcile the two conflicting descriptions of chiral stability, the
classical ``localized'' one and the quantum ``delocalized'' one
\cite{qzeno0,decoherence_review}.
To invoke a two-level approximation for the system, we assumed that
the potential barrier is very high,  or equivalently the energy of
the system is sufficiently low that the higher energy levels are not
involved in the dynamics. In addition, since we introduced the
environment, we should point out that the position measurement of
the system is known to increase both its energy $\langle \hat H
\rangle$ and energy width $\delta E\equiv \sqrt {\langle
 \hat H^2\rangle-\langle  \hat H \rangle ^2}$, where $\langle\cdot\rangle$
denotes an ensemble-average, over time \cite{hhan_jcp, gallis}. The
system energy and energy width will eventually become greater than
the next higher energy levels, leading to the breakdown in the
validity of the two-level approximation. The stronger
$\gamma^{\prime}$, the shorter the time during which the two-level
approximation will be valid.
Using the two-level approximation and the evenness of $\k{+}$ and
oddness of $\k{-}$, we derive the position measurement operator as
\begin{eqnarray}
\hat{x} & \approx & \k{+}\b{+}\hat{x}\k{-}\b{-}+{\mrm{H.c}}\nonumber \\
 & = & x_{\mrm{avg}}\hat{\sigma_{z}}
 \approx  x_{\mrm{min}}\hat{\sigma_{z}},\label{x}\end{eqnarray}
where ${\mrm{H.c}}$ denotes the Hermitian conjugate and
$\hat{\sigma_{z}}\equiv(\k R \b R -\k L \b L)$.  Here $\pm
 x_{\mrm{min}}$
are the positions of the double-well minima, which for a deep well
are very close to the average positions of the right and left wave
functions (denoted by $\pm x_{\mrm{avg}}$ with definitions $
x_{\mrm{avg}}\equiv \b R \hat{x}\k R$ and $ -x_{\mrm{avg}}\equiv
\b L \hat{x}\k L$). Then the master equation for the two-level
system becomes \cite{deco_model}
\begin{eqnarray}
\PD{\hat{\rho}}{t}=\frac{i}{\hbar}[\HAH,\hat{\rho}]-\gamma[\hat{\sigma}_{z},[\hat{\sigma}_{z},\hat{\rho}]],\label{liouville12l}\end{eqnarray}
with the dephasing rate $\gamma\equiv\gamma^{\prime}
x_{\mrm{min}}^{2}$. This equation gives the evolution of the
two-level system subjected to $\hat{\sigma}_{z}$ representing a
continual measurement of the system coordinate $\hat{x}$. In the
first term we will use the Hamiltonian of the two-level system with
the intermolecular interactions under a mean-field approximation as
used in Eqs. (\ref{nonlinearEL}) and (\ref{nonlinearER}).  The
second term in this equation represents a measurement of a
population difference between two states $\k{L}$ and $\k{R}$, and
leads to decay of the coherences between the $\k{L}$ and $\k{R}$
states (decay of the off-diagonal matrix elements, $\rho_{LR}$).
To amalgamate Eqs. (\ref{nonlinearEL}), (\ref{nonlinearER}) and
Eq. (\ref{liouville12l}), we rewrite Eq.   (\ref{liouville12l})
into the stochastic  Schr\"{o}dinger equation for the state vector
$\k{\Psi}$ \cite{perci1,deco_model}:
\begin{eqnarray}
\k{d\Psi}=\left(-\frac{i}{\hbar}\HAH-\gamma\left[\hat{\sigma}_{z}-\langle\hat{\sigma}_{z}\rangle_{\k{\Psi}}\right]^{2}\right)\k{\Psi}dt+\sqrt{\gamma}\left[\hat{\sigma}_{z}-\langle\hat{\sigma}_{z}\rangle_{\k{\Psi}}\right]\k{\Psi}d\xi,\label{QSD}
\end{eqnarray}
where $\langle\hat{\sigma}_{z}\rangle_{\k{\Psi}}=\b{\Psi}\hat{\sigma}_{z}\k{\Psi}$
and $\xi$ is for a complex Wiener process \cite{perci1}. Combining
the intermolecular interactions and the decoherence leads to the
resultant Schr\"{o}dinger equation in terms of $a_{L}$ and $a_{R}$,
which is stochastic and nonlinear: \begin{eqnarray}
\dot{a_{L}} & = & -\frac{i}{\hbar}\left(E_{m}a_{L}-\delta a_{R}\right)-\frac{i}{\hbar}\left(V_{hom}|a_{L}|^{2}+V_{het}|a_{R}|^{2}\right)a_{L}-\left(\gamma|a_{R}|^{4}+\sqrt{\gamma}\dot{\xi}(t)|a_{R}|^{2}\right)a_{L},\label{nonlinearEL2}\\
\dot{a_{R}} & = & -\frac{i}{\hbar}\left(E_{m}a_{R}-\delta
a_{L}\right)-\frac{i}{\hbar}\left(V_{hom}|a_{R}|^{2}+V_{het}|a_{L}|^{2}\right)a_{R}-\left(\gamma|a_{L}|^{4}-\sqrt{\gamma}\dot{\xi}(t)|a_{L}|^{2}\right)a_{R}.\label{nonlinearER2}\end{eqnarray}
Here, the time-evolution is governed by three components. The first
and the second components are responsible for  the dynamics of the
isolated system and for  the intermolecular (hetero/homochiral)
interactions between the single molecule with the other surrounding
molecules within the mean-field approximation \cite{vardi},
respectively. The last component models the interaction with the
environment that induces the loss of quantum coherence in the
system. Noting that $|a_R|^2+|a_L|^2=1$, and rescaling the time as
$\omega t\rightarrow t$, Eqs. (\ref{nonlinearEL2}) and
(\ref{nonlinearER2}) become
\begin{eqnarray}
 \dot{a_L}& =& - i \left(\Omega+v|a_R|^2\right)a_L  +
 i a_R
 -  \left(\Gamma |a_R|^4
+\sqrt{\Gamma}\dot{\eta}(t)|a_R|^2\right) a_L,\label{nonlinearEL3}\\
 \dot{a_R}& =& - i \left(\Omega+v|a_L|^2\right)a_R    + i a_L
 -  \left(\Gamma |a_L|^4
-\sqrt{\Gamma}\dot{\eta}(t)|a_L|^2\right) a_R, \label{nonlinearER3}
\end{eqnarray}
with $\Omega=(E_m+V_{hom})/\delta$, $v=(V_{het}-V_{hom})/\delta$,
$\Gamma=\gamma/\omega$ and $d\eta=\sqrt{\omega}d\xi$.
We gradually build up a more complicated system starting from a
simple system. In section \ref{deco_only} we study the system in
the presence of decoherence effect only, using   Eq.
(\ref{liouville12l}) with $\hat{H}=\hat{H}_0$ where $\hat{H_0}$ is
given in Eq.(\ref{singleH}). In section \ref{inter_only} we study
the system in the presence of intermolecular interactions only,
using Eqs.(\ref{nonlinearEL}) and (\ref{nonlinearER}). In section
\ref{both}, we study the system in the presence of  the combined
effects of the decoherence and the intermolecular interactions,
using Eqs.(\ref{nonlinearEL3}) and (\ref{nonlinearER3}).
\section{the model}\label{model}
\subsection{Initial condition: Localized and delocalized states}\label{conditions}
We examine three cases for the pure state with an initially
different coherence, or equivalently with an initially different
degree of localization. For an initially localized state (denoted as
LS), we choose $\k{R}$, located in the right-well. For an initially
delocalized state, we examine two cases, a weakly delocalized state
(denoted as WDS) and a strongly delocalized state (denoted as SDS),
with $\sqrt{0.05}\k{L}+\sqrt{0.95}\k{R}$ and
$\sqrt{0.49}\k{L}+\sqrt{0.51}\k{R}$, respectively. For the LS case,
the initial coherence ($\equiv\rho_{LR}$) is 0, and the initial
population difference between the two chiral states, $Z$ ($\equiv
\rho_{RR}-\rho_{LL}$) is 1. For the WDS case, initially $\rho_{LR}$=
$\sqrt{0.05\times0.95}\approx0.2$, and $Z$=0.9. For the SDS case,
initially $\rho_{LR}$= $\sqrt{0.49\times0.51}\approx0.5$, and
$Z$=0.02.
\subsection{Measures for the distinguishability}\label{measure_distinguishability}
With the respect to the question of whether one can distinguish a
pure state from a corresponding mixture with the same populations
in the right and left wells. Harris and coworkers
\cite{harris_distinguish} showed, in the framework of the
wavefunction, that no parity sensitive experiment can measure the
difference between them if the states $\k L $ and $\k R$ are taken
to be real. They assumed that instantaneous measurements occur on a
time scale short compared to that for the system dynamics (tunneling)
and compared to that for any interactions with the surrounding
molecules. On the contrary, we include an external environment that
destroys the right-left coherences in the framework of the
density-matrix, and the intermolecular interactions (the homochiral
and heterochiral interactions) in the framework of the wavefunction
within a mean-field theory.

We investigate the distinguishability between the initially pure and mixed states by measuring a population in
the right- well.   For the measure, we compare $\rho_{RR}$ of the initially pure vs. mixed state, and for the system with decoherence only, we use $\Delta Z (\equiv Z^{\mrm{P}}- Z^{\mrm{M}})$ where superscripts P and M stand for the initially pure state and the corresponding mixed state.

To examine the distinguishability between the initially pure state
and the corresponding mixed state over time, we are required to
select the initial conditions for the pure state
$(X_0^{\mrm{P}},Y_0^{\mrm{P}},Z_0^{\mrm{P}})$ and the mixed state
$(X_0^{\mrm{M}},Y_0^{\mrm{M}},Z_0^{\mrm{M}})$ where $X\equiv
2{\mrm{Re}}\,(\rho_{LR})$,  $Y\equiv 2{\mrm{Im}}\,(\rho_{LR})$, and
($X, Y$,  $Z$)=($X_0, Y_0$,  $Z_0$) at $t=0$. For the maximum
contrast between the pure and the mixed state, we choose the same
initial population of $\k L$ and $\k R$ in the initial pure and
mixed states, \ie $Z_0^{\mrm{P}}=Z_0^{\mrm{M}}$, but choose  the
degree of initial coherence to be zero for the mixed state (maximum
randomness), \ie $X_0^{\mrm{M}}=Y_0^{\mrm{M}}=0$, while varying
initial coherences for the pure state. In other words, when the
initially pure state is given by \begin{eqnarray}\k
{\Psi(t=0)}=a_{{L}0}\k L +a_{{R}0}\k
R,\label{psi0pure}\end{eqnarray} or, equivalently,
\begin{eqnarray}\hat{\rho}^{\mrm{P}}(t=0)=|a_{{L}0}|^2\k L\b L
+|a_{{R}0}|^2\k R\b R+a_{{L}0}a_{{R}0}^{*}\k L\b
R+a_{{R}0}a_{{L}0}^{*}\k R\b L,\label{rho0pure}\end{eqnarray} the
corresponding mixed state is given by
\begin{eqnarray}\hat{\rho}^{\mrm{M}}(t=0)=|a_{{L}0}|^2\k L\b L
+|a_{{R}0}|^2\k R\b R.\label{rho0mixed}\end{eqnarray} This initially
mixed state may be prepared by making the system interact with an
environment that extinguishes the left-right coherence of the system
sufficiently fast so that the system dynamics is ignorable.
\subsection{Measures for the degree of decoherence: Purity and
off-diagonal elements}\label{measures}
In order to measure coherence decay, \ie, decoherence, we use both
the purity and the off-diagonal elements of the reduced density
matrix. The purity $\varsigma$ is defined as
\cite{purity}\begin{eqnarray}
\varsigma\equiv\mathrm{Tr}(\hat{{\rho}}^{2}),\label{purity}\end{eqnarray}
 where $\mathrm{Tr}$ denotes a trace over the system of interest.
Note that the decoherence process does not preserve the purity of
the state, that is, the purity of the resultant mixed state becomes
less than 1, while that of the pure state is 1.  In particular,  for
a two-level system purity becomes, in the $\k{L}$, $\k{R}$ basis,
\begin{eqnarray}\mathrm{Tr}(\hat{{\rho}}^{2})=\rho_{LL}^{2}+2|\rho_{LR}|^{2}+\rho_{RR}^{2},\label{purity_LR}\end{eqnarray}
and is given by, in the Bloch-vector representation,
\begin{eqnarray}\varsigma=\frac{1}{2}(X^2+Y^2+Z^2+1).\label{puritydef2}\end{eqnarray}
As can be easily seen in Eq. (\ref{puritydef2}), $1/2\le\varsigma\le
1$.  When the purity decays to a minimum value, 1/2, this
corresponds to $X=Y=Z=0$, which represents a fully mixed state (a
racemic mixture), with equal populations in the right and left wells
and with no remaining coherence between them.  On the other hand,
when the purity remains close to 1, it implies that the system is
resistent to the decoherence process.
Another more detailed, albeit representation-dependent, measure of
decoherence is provided by the decay of off-diagonal reduced
density matrix elements. The diagonal elements are the
probabilities of finding the system in the corresponding basis
state. The off-diagonal elements are a measure of coherence.
Specifically, in the $\k{L}$, $\k{R}$ basis, $\rho_{LL}$,
$\rho_{RR}$ are the probabilities of finding the system in the
state $\k{L}$ (left well) and $\k{R}$ (right well), respectively,
and $\rho_{LR}$ is the coherence between the two states.
When numerically examining the decoherence rate in terms of the
purity decay or the decay of off-diagonal elements,  the initially
pure states are chosen, rather than the initially mixed states, in
order to maximize the influence of initial coherences.  That is,
$\varsigma(t=0)$=1 ($X_0^2+Y_0^2+Z_0^2=1$).
\subsection{Computational methods}\label{computation}
For the system with decoeherence only, we obtained an analytical
solution \cite{cplhhan_wardlaw}. However, for the system with
intermolecular interactions only and the system with both the
intermolecular interactions and the decoherence effect, we had to
employ computational methods. The terms involved with the stochastic
terms $\eta$ in Eqs.(\ref{nonlinearEL3}) and (\ref{nonlinearER3}) are
integrated by a first order scheme, and the other terms are
integrated by the fourth-order Runge-Kutta method. The values of
$a_{L}$ and $a_{R}$ resulting from the stochastic Schr\"{o}dinger
equation
(\ref{nonlinearEL3}) and (\ref{nonlinearER3}), when averaged over
many realizations of the Wiener process, provide the density matrix
elements, probabilities and coherences, according to the following
relations: $\b L \hat{\rho}\k L=\rho_{LL}=\langle
a_{L}a_{L}^{*}\rangle$, $\b R \hat{\rho}\k R=\rho_{RR}=\langle
a_{R}a_{R}^{*}\rangle$, and $\b L \hat{\rho}\k R=\rho_{LR}=\langle
a_{L}a_{R}^{*}\rangle$. A time step of $10^{-3}$ to $10^{-4}$ and
an ensemble of $10^4$ realizations were used.
\section{system with decoherence only}\label{deco_only}
To provide a simple physical picture focused on the decoherence
effect in affecting a tunneling motion between the two wells, we
consider a system with decoherence only, i.e., there are no
intermolecular interactions and the chiral discrimination
energies are zero in Eqs. (\ref{nonlinearEL3}) and
(\ref{nonlinearER3}) ($v=0$). In our work \cite{cplhhan_wardlaw}, we
study the roles of initial coherences of the system and dephasing
rates on the distinguishability of initially pure versus initially
mixed states, and the vulnerability to decoherence, \ie, how fast the system will lose its coherence or become a randomly mixed state. In this section
we will briefly review the main results of Ref. \cite{cplhhan_wardlaw}
and then further discuss the effect of the tunneling frequency $\omega$
on the distinguishability and the purity decay rate.
\subsection{Review: Effect of initial coherence and dephasing rate}
Within a two-level approximation, using $\hat{H}=\hat{H_0}$,
for simplicity choosing the energy origin to correspond to
$E_m=0$, \ie, $E_+=-E_-$, and switching from the density-matrix
elements of the states $\k L$ and $\k R$, $\rho_{LL}$, $\rho_{RR}$,
and  $\rho_{LR}$ to the real quantities, $X$, $Y$, and $Z$ for the
Bloch-vector representation, Eq. (\ref{liouville12l}) gives the
following set of equations (see, for details,
Ref. \cite{cplhhan_wardlaw}) :
\begin{eqnarray}
\frac{dX}{dt}&=&-4\gamma X,\label{X}\\
\frac{dY}{dt}&=&-4\gamma Y +2\omega Z,\label{Y}\\
\frac{dZ}{dt}&=&-2\omega Y.\label{Z}\end{eqnarray} The resultant
equations are of standard form \cite{torrey}. At long times, when
the system reaches equilibrium, $dX/dt=dY/dt=dZ/dt=0$, and the
steady-state solution to Eqs. (\ref{X})-(\ref{Z}) becomes $X=Y=Z=0$.
This stationary state represents a fully mixed state, with equal
populations in the right and left wells and with no remaining
coherence between them, and  is reached regardless of the initial
conditions, the system's tunneling frequency  $\omega$, and the
dephasing rate $\gamma$ (related to the nature of the environment
and the potential of the system).
With an arbitrary initial state {[}$X=X_{0},Y=Y_{0},Z=Z_{0}${]}, the
general analytical solution (not shown here) can be obtained
separately for $\gamma<\omega$ (tunneling dominant region) and for
$\gamma>\omega$ (dephasing dominant region). The general solution
reveals that, in the dephasing dominant regime for an initially
localized state, say $\k R$ (note that this corresponds to an
initially chiral state), the system approaches the steady state
(fully mixed state) monotonically without any oscillations present
but, however, as $\gamma$ increases, tends to be stuck in the right
well on a time scale of $1/\gamma$ (this behavior is equivalent to the
quantum Zeno effect \cite{deco_model}). Ref. \cite{cplhhan_wardlaw}
demonstrated that, as $\gamma$ increases in this dephasing dominant
region, an initially chiral system  tends to preserve the chirality,
while an initially very delocalized state such as the SDS case tends
to become a racemic mixture more quickly.
From a general solution to Eqs.(\ref{X})-(\ref{Z}), the difference
between the time-evolution of $Z$ of the initially pure state and
that of the corresponding mixed state, $\Delta Z (\equiv Z^{\mrm{P}}-
Z^{\mrm{M}}) $ is given by \cite{cplhhan_wardlaw}:
\begin{eqnarray}
\Delta Z(t)= -\frac{2\omega Y_0^{\mrm{P}}
 }{s} e^{-2\gamma t}\sin (st)\,\, &{\mrm{for}}&
\omega>\gamma,\label{z_diff1}\\
\Delta Z(t) = -\frac{\omega Y_0^{\mrm{P}} }{\tilde{s}}
[e^{(\tilde{s}-2\gamma )t}-e^{-(\tilde{s}+2\gamma )t}]\,\,
&{\mrm{for}}& \omega<\gamma \label{z_diff2},
\end{eqnarray}  with $s=2\sqrt{\omega^2-\gamma^2}$ and $\tilde{s} = 2\sqrt{\gamma^2-\omega^2}$.
At a glance, one can easily see that the value of $Y_0^{\mrm{P}}$ is
crucial to the distinguishability. If $Y_0^{\mrm{P}}=0$, $\Delta
Z(t)=0$, which means no distinguishability, for all time and for all
$\gamma$. However, if $Y_0^{\mrm{P}}\neq 0$ the distinguishability
does exist, although it disappears over time because the decoherence
impels both the initially pure state and the corresponding mixture
to the maximally randomized stationary state.
For the tunneling dominant region, as $\gamma$ increases, the period
of the oscillation increases and the amplitude decays faster, leading
to $\Delta Z$ approaching zero more rapidly, at which point one can
not tell whether the system was initially in the pure state or in the
corresponding mixture upon the measurement of population. However, for
the dephasing dominant region, as $\gamma$ increases, the
distinguishability disappears more slowly, suggesting that considerably
strong dephasing may unexpectedly afford means to observe the long-time
surviving distinguishability.
Noting that for a two-level system purity is given by
Eq.(\ref{puritydef2}) and using Eqs.(\ref{X})-(\ref{Z}), one can
obtain
\begin{eqnarray}
 \frac{d\varsigma}{dt}=-4\gamma(X^2+Y^2).\label{puritydt}
\end{eqnarray}
In our previous work \cite{cplhhan_wardlaw}, the interplay of the
initial coherence and the dephasing rate on the purity decay, in
particular, as it relates to the racemization and chirality
stabilization, was clarified. For very short times, increasing the
dephasing rate $\gamma$ and the initial coherences
($X_0^2+Y_0^2=4|\rho_{LR}|^2$) increases the purity decay
(decoherence) rate. On the other hand, for later times, for larger
initial coherences ($X_0^2+Y_0^2\approx 1$), the purity still tends
to decay faster with increasing $\gamma$, but for the smaller
initial coherences ($X_0^2+Y_0^2\approx 0$), the purity shows the
opposite behavior when $\gamma$ is increased beyond a certain
threshold, \ie, the purity decay starts to be suppressed. This
interesting observation enables us to suggest that, in the
environment destroying the left-right coherences,  on a timescale
faster than the intra-system dynamics (the tunneling time scale),
the initially strongly localized state would tend to preserve the
chirality, while, on the other hand, an initially strongly
delocalized state such as a superposition of the two chiral states,
would become a racemic mixture  very quickly.
It is notable  that the observation agrees with the prediction that
if the system dynamics is ignorable compared with the interaction
between the system and the environment, the coherence between the
eigenstates of the decohering operator will be lost fastest
\cite{qzeno0,decoherence_review}.  The applicability of the
prediction can be seen by realizing that the two chiral states $\k
L$ and $\k R$ correspond to the eigenstates of the decohering
operator ($\hat\sigma_{z}=\k R \b R - \k L \b L$), and the stronger
dephasing rate beyond a certain threshold allows us to ignore the
system dynamics compared with the decoherence process. Then,
according to the prediction, a superposition of the two chiral
states will lose its coherence much faster than a localized chiral
state ($\k L$ or $\k R$). However, it is also notable that there was
previously  no prediction for the observation in Ref.
\cite{cplhhan_wardlaw} that increasing the initial coherences
($X_0^2+Y_0^2$) leads to a system that exhibits more rapid
decoherence.
\subsection{Tunneling frequency $\omega$ (=$\delta/\hbar$)}
The typical behavior of $\Delta Z(t)$ for several values of $\omega$
for a given value of $\gamma$ (here chosen as 1) is shown in Fig.
\ref{newddeltaz} (a) and (b). For $\omega<\gamma $ (dephasing dominant
region, here $\omega < $ 1, shown in (a)), as time increases $\Delta Z
/Y_0^{\mrm P}$ first decreases from zero to a minimum and then increases
monotonically towards zero. Note that, as $\omega$ increases, the
maximum of $|\Delta Z /Y_0^{\mrm P}|$ gets larger and is obtained at an
earlier time, \ie, $t_{\mrm{max}}$ gets smaller and the
distinguishability disappears more rapidly beyond $t_{max}$ (i.e.
$\Delta Z /Y_0^{\mrm P}$ goes to zero more quickly). For instance,
compare the cases $\omega=0.1$ and $\omega=0.8$ in Fig. \ref{newddeltaz}
(a). For the case $\omega=0.1$, $|\Delta Z /Y_0^{\mrm P}|=0.0494$
at $t=t_{\mrm{max}}\approx 1.5$, and, is still nonzero at $t = 8$
having a value of 0.0464 which is roughly the same as the maximum
value of $|\Delta Z /Y_0^{\mrm P}|$), indicating the long-time
surviving distinguishability, while for the case  $\omega=0.8$,
$|\Delta Z /Y_0^{\mrm P}|=0.315$ at $t=t_{\mrm{max}}=0.58$, and is
almost zero at $t=8$, indicating no distinguishability for $t>8$.
This observation suggests that a very slow tunneling
oscillation may unexpectedly allow us to observe the long-time
surviving distinguishability. On the other hand, for $\omega >
\gamma$ (tunneling dominant region, here $\omega > $ 1, shown in
(b)), as $\omega$ increases, the period of the oscillation in
$\Delta Z$ decreases, but the amplitude envelope decays exponentially
with a rate that is almost independent of $\omega$. In other words,
for the tunneling dominant region, the rate with which $\Delta Z$
approaches toward zero is largely determined by $\gamma$, as expected
from Eq.(\ref{z_diff1}). For all the cases $\omega=2,4,6$
[Fig. \ref{newddeltaz} (b)], upon the measurement of population after
$t=3$, one can not tell whether the system was initially in the pure
state or in the corresponding mixture.
Figure \ref{newddeltaz}(c) captures well the effect of increasing
the tunneling frequency of the system on the purity decay at a given
$\gamma$ (= 1). Initially the system was prepared in the $\k R$ state
(the LS case). Interestingly and non-intuitively, increasing the
tunneling frequency $\omega$ tends to increase the rate of purity
decay  rather than reducing it. For the case $\omega$=0.2, 0.4 and 0.8
(the dephasing dominant region), increasing $\omega$ clearly
accelerates the purity decay for all time. Even for the case $\omega$ =
2, 4, and 8 (the tunneling dominant region) where the oscillatory decay
becomes visible, the purity decay rate increases with increasing
$\omega$ at short times, while the overall decay timescale appears to
be dominated by the dephasing rate $\gamma$.
In regard to the quantum-classical transition, realizing that a
smaller $\omega$ corresponds to a higher inter-well barrier, a
longer tunneling time, and thus a more classical system, the
observation that decreasing $\omega$ decreases the decoherence rate
can be put in the following way; a more classical (quantum) system
is less (more) vulnerable to decoherence.
It is also noteworthy that the master equation of  the system of
interest in this section (the chiral system within a two-level
approximation in the presence of an environment destroying the
coherence between the two states, see Eqs. (\ref{X}) to
(\ref{Z})) is identical to that resulting from a two-level atom
driven by a field undergoing a pure dephasing \cite{qoptics} in
quantum optics, as previously mentioned in Ref. \cite{deco_model}.
This equivalency emerges where the following replacements are made:
$\k {L,R}\rightarrow$ the two energy eigenstates, $\omega
\rightarrow$ Rabi-frequency, and $\gamma\rightarrow$ pure dephasing
rate. Noting this equivalency enables us to recognize that
the above observation (increasing $\omega$ tends to increase the
purity decay rate) is also equivalent to the result of
Ref. \cite{sanz_brumer}, where aspects of coherence and decoherence
were studied with a two-level system interacting with a decohering
environment and a resonant continuous-wave electromagnetic field, that
increasing the Rabi frequency increases the decoherence rate.
\section{system with intermolecular interactions
only}\label{inter_only}
The system with intermolecular interactions
within a mean-field approximation was studied in Ref. \cite{vardi},
with a particular focus on the role of intermolecular interactions in
establishing chiral stability.  Localized chiral states were shown
to be stable when the homochiral interactions are energetically
favorable to heterochiral interactions ($v>0$), whereas delocalized
states are dominant when the heterochiral interactions are
energetically favorable ($v<0$). For a self-completeness and a later
comparison with the case in the presence of both the intermolecular
interactions and the decoherence effect, we reproduce some of the
results of Ref. \cite{vardi}, with emphasis on the interplay of the
chiral discrimination energy parameter $v$ and the initial condition
of the system in establishing the anharmonic oscillations in population
of the two wells of the system. In subsection
\ref{distinguishability_vardi} we show that a nonzero chiral
discrimination energy can bring about the distinguishability between
the initially pure and the corresponding mixed states even for the
$Y_0^{\mrm{P}}=0$ case, where there is no distinguishability in the
system in the absence of the intermolecular interactions ($v=0$),
regardless of the decoherence effect, as shown subsection
\ref{deco_only}.
\subsection{Review}\label{review}
The system with intermolecular interactions only is described by
Eqs.(\ref{nonlinearEL3}) and (\ref{nonlinearER3}) with $\Gamma=0$.
Below for all the numerical calculations $\Omega$ is chosen as 1.
The time-evolution of the initially localized state $\k R$ (the LS
case) with a variation of $v$ is shown in Fig.\ref{figg001}. For
the $v=0$ case we see coherent tunneling from one well to the other.
As $|v|$ increases (the chiral energy difference $|V_{het}-V_{hom}|$
increases), the nonlinearity of the differential equations increases
and the oscillations in population between the two wells becomes
anharmonic. For a sufficiently large $|v|(> 4)$ and with increasing
$|v|$, population-trapping becomes stronger and maximum coherence
decreases. We observed a similar relation between the coherence
$|\rho_{LR}|^2$ and the self-trapping in the WDS and the SDS cases
(not shown here): when self-trapping due to intermolecular interactions
is manifested in $\rho_{RR}$, $|\rho_{LR}|^2$ is seen to exhibit a
decrease in the oscillation amplitude, and thus a decrease in the
maximum coherence, which prohibits a complete tunneling between the
two wells, and consequently results in the self-trapping of population.
Considering that characteristic values of the chiral discrimination
energy difference, $|V_{het}-V_{hom}|$, are far greater than the
tunneling splitting $\delta$, \ie, $|v|\gg 1$ \cite{alkorta},
Fig. \ref{figg001}(a) demonstrates that much of the system in a
condensed phase can be self-trapped in one of the enantiometric forms
to a degree which depends on the chiral discrimination energies (i.e.
$V_{het}$ and $V_{hom}$).
As demonstrated in Fig. \ref{figg001}, behaviors of $\rho_{RR},
|\rho_{LR}|^2$, and thus the self-trapping effect are independent of
the sign of $|v|$ for the initially localized state (LS case). Note
that the LS case has no initial coherence between $\k L$ and $\k R$,
\ie, $\rho_{LR}(t=0)=0$. However, when there exist nonzero initial
coherences, the stability of the initial states does depend on the
sign of $|v|$, \ie, whether the homochiral mixture is energetically
favorable ($v>0$) or the heterochiral mixture is favorable ($v<0$).
We present two cases of different initial coherences, the WDS and
SDS cases, for several values of $v$. For the WDS case with positive
$v$ [Fig. \ref{figg00595}(a)], as $v$ increases from 0 to 2 the
oscillation period of $\rho_{RR}$ increases and then decreases beyond
$v = 2$ with the population becoming self-trapped for $v \geq 5$.
This observation is similar to that of the LS case. However, when $v$
is negative [Fig. \ref{figg00595}(b)], this initially small degree of
delocalization leads the system to oscillate between the two wells
(chiral activity) with a faster oscillation frequency than that of
the $v=0$ case, until $v$ reaches quite a large magnitude, for instance,
$v=-10$, when the system is now self-trapped again. Comparisons of the
$v=5$ and the $v=-5$ cases show how the initially weakly delocalized
state (the WDS case) can behave very differently depending on the sign
of $v$. For large negative $v$ (e.g., $v = - 10$) the system becomes
self-trapped. For large positive $v$ (e.g., $v = + 5$) the system also
becomes self-trapped, but such a trapping is stable in sense that it
remains self-trapped as $v$ is further increased.
For positive $v$ in the WDS case the self-trapping becomes very stable,
while for the negative $v$ the tunneling is more assisted and stabilized
by the addition of the intermolecular interactions. This behavior is
shown in the exactly opposite way for the SDS case, since the initial
conditions with almost equal populations in each well ($a_{{L}}\approx
a_{{R}}$) correspond to stable states for negative $v$ but unstable
states for positive $v$. With positive $v$ [Fig. \ref{figg04951}(a)] the
initially strongly delocalized state exhibits oscillations about
$\rho_{RR}\approx$ 0.75 for $v=5$ and $\rho_{RR}\approx$ 0.7 for $v=10$,
indicating that the system is chiral active, whereas as for $v = 0$ and
$v = 2$, $\rho_{RR}$ exhibits a small amplitude oscillations about $\sim
0.5$, indicating that the system is not chilar active. On the other hand,
for negative $v$ [Fig. \ref{figg04951}(b)] the delocalized state becomes
very stable around $\rho_{RR}=0.5$ with an oscillation frequency that
increases with increasing $|v|$.
\subsection{Distinguishability of pure and mixed
states}\label{distinguishability_vardi}
One of the main motivations of this paper was to answer some questions
arising from Vardi's claim in the last paragraph of Ref.
\cite{vardi}.  He argued that from the results ``{\it{$\cdots$ In the
former case, spontaneous symmetry breaking may take
place, turning a nearly racemic mixture into an optically active
ensemble}}.''  Since his model does not associate with any
decoherence source, the system remains pure if it is initially pure.
All of his results were obtained for initially pure states. Thus
when he used ``racemic mixture'', it would have been correctly
described as a delocalized pure state as opposed to a racemic mixture
that is found in nature and associated with loss of coherences. At
this point we face two questions. First, if the system starts as a
mixed state, then will it show results similar to those of initially
pure states? This question may be put as follows: can we distinguish
the initially pure and the initially mixed states for the system with
intermolecular interactions? If we cannot do so by measuring
populations, for instance, $\rho_{RR}$, then Vardi's claim may be
justifiable. Otherwise, decoherence should be seriously considered
to answer the second question: how can an initially pure state end up
as a nearly racemic mixture or an optically active ensemble in real
system?  We attempt to answer the first question here and the second
one in the next section.
We start by showing the time-evolutions of population and coherence in
Fig. \ref{figg0v0mp} with the initially pure and
the corresponding mixed state for the WDS and SDS cases (note that
for both of these cases $Y_0^{\mrm{P}}=0$) in the absence of the
intermolecular interactions ($v=0$).    Figure \ref{figg0v0mp}(a)
demonstrates that when $Y_0^{\mrm{P}}=0$, there is no
distinguishability in terms of the population ($\Delta Z=0$), as
already seen in Eqs. (\ref{z_diff1}) and (\ref{z_diff2}), whereas
there are clear differences in terms of the coherence as shown in
Fig. \ref{figg0v0mp}(b). Now we proceed the issue of
distinguishability of the initially pure and mixed states for the
system with the intermolecular interactions. In the previous section
\ref{review}, changing the initial condition from the LS case to the
WDS and SDS cases, and thus creating nonzero initial coherence is
seen to lead the system to behave differently depending on the sign
of $v$.  Thus, one may imagine that the initially mixed state
without initial coherences may behave differently from the initially
pure state with nonzero initial coherence, even when
$Y_0^{\mrm{P}}=0$ such as in the WDS and the SDS cases. We compare
the initially pure state for the WDS and SDS with their
corresponding mixed states, in terms of the population and the
coherence, in particular, for $v=5$ and $v=-5$. Figure
\ref{figg0v0mp}(c) demonstrates that for the WDS case  a nonzero
$v$ enables one to tell whether the system was initially in a pure
or a mixed state by measuring populations. The pure state is
observed to behave differently, depending on the sign of $v$, while
the behavior of the corresponding mixed state is independent of the
sign of $v$ due to the absence of its initial coherence. Since the
WDS case is slightly delocalized from the LS case, the behavior of
the population (as measured by $\rho_{RR}$) of the initially mixed
state for the WDS case is quite close to that of the pure state of
the LS case [Fig. \ref{figg0v0mp}(c)]. For the SDS case, the similar
distinguishability, brought about by the nonzero $v$, is seen in
Fig. \ref{figg0v0mp}(d). Let us clarify the behavior of the initially
mixed state  related to the LS case with the same $v(=\pm 5)$: one
can note that $\rho_{RR}$ of the mixed state for the WDS and SDS
cases is seen to oscillate in the same way as that of the initially
pure state for the LS case with the same $v$ and differences are
found only in the amplitude and the associated mean value $\rho_{RR}$
(as obtained by averaging over an oscillation cycle) [Figs.
\ref{figg0v0mp}(c) and \ref{figg0v0mp}(d)]. When the initially pure
state is given by Eq. (\ref{psi0pure}), the corresponding mixed state
is chosen by Eq. (\ref{rho0mixed}). Then, the probability of being in
the right well for the initially mixed state, $\rho^{\mrm{M}}_{RR}(t)
= \b{R}\hat{\rho}^{\mrm{M}}(t)\k{R}$ can be expressed in terms of the
probability of being in the right well for the LS case (denoted as
$|a_{{R}}^{RR}(t)|^2$):
\begin{eqnarray}
\rho^{\mrm{M}}_{RR}(t)&=&\langle |a_{{R}}|^2\rangle,\\
&=&
|a_{{L0}}|^2|a_{{R}}^{LL}(t)|^2+|a_{{R0}}|^2|a_{{R}}^{RR}(t)|^2,\\
&=&|a_{{L0}}|^2+(|a_{{R0}}|^2-|a_{{L0}}|^2)|a_{{R}}^{RR}(t)|^2.
\label{rhoRRmixed}
\end{eqnarray}
where $|a_{{R}}^{LL}(t)|^2$ ( $|a_{{R}}^{RR}(t)|^2$) denotes the
probability of being in the right well at time $t$ for the state
initially localized in $\k L$ ($\k R$). Here we used
$|a_{{R}}^{LL}(t)|^2=1-|a_{{L}}^{LL}(t)|^2=1-|a_{{R}}^{RR}(t)|^2$:
the second equality holds since Eqs.  (\ref{nonlinearEL3}) and
(\ref{nonlinearER3}) with $\Gamma=0$ (decoherence-free case) are L-R
symmetric.
The above Eq. (\ref{rhoRRmixed}) indicates that the probability of the
initially mixed state should have the same oscillation period as
that of the initially pure state  localized in $\k R$ (LS case) with
the same $v$, but the amplitude and  average of the probability of
the former are changed from those of the latter, depending on its
$|a_{{L0}}|^2$, and $|a_{{R0}}|^2$.  For the SDS case, for instance,
since $|a_{{L0}}|^2=0.49$, and $|a_{{R0}}|^2=0.51$,
$\rho^{\mrm{M}}_{RR}$ of the initially mixed state with $v=\pm5$
oscillates from 0.51 to 0.506 in  tune with $\rho^{\mrm{P}}_{RR}$
of the initially pure state localized in  $\k R$ (LS case)
($=|a_{{R}}^{RR}(t)|^2$) with the same $v(=\pm5)$ which oscillates
from 1 to 0.8 [Fig. \ref{figg0v0mp}(d)]. The similar observation is
made in the WDS case as well [Fig. \ref{figg0v0mp}(c)].
As clearly demonstrated in Figs. \ref{figg0v0mp}(c) and
\ref{figg0v0mp}(d), for a system with intermolecular
interactions, the initially mixed state behaves differently from the
initially pure state as measured by the probabilities. In addition,
the initially mixed state does not behave differently depending on
the sign of $v$. Thus, Vardi's claim is unjustifiable. For instance,
an initially nearly racemic mixture  remains as a racemic mixture,
regardless of whether the chiral discrimination energy favors the
homochiral mixture or the heterochiral mixture, as shown in
Fig. \ref{figg0v0mp}(d). Now it seems very critical to include the
initial coherences adequately to obtain the correct behavior of the
system subject to the intermolecular interactions. Further, if one
were to create a nearly racemic mixture or an optically active
ensemble in laboratory, starting from an initially pure state, it
would be essential to include the decoherence process in the
assessment and modeling of the system dynamics. We explore this in
the following section.
\section{system in the presence of both decoherence and intermolecular
interactions} \label{both}
In this section, we finally consider the system in the
presence of both a decoherence effect and intermolecular
interactions. In Section \ref{deco_only} we discussed general
features of the decoherence effect on the system without chiral
discrimination energies, \ie, $v$=0. Two prominent system behaviours
were observed: suppressing the tunnelings (oscillations) between the
two chiral states by destroying the coherence, and forcing the system
to the long-time mixed state with a complete randomness,
$\hat\rho(t\rightarrow\infty)=\frac{1}{2} (\k L \b L+\k R\b R)$,
regardless of the dephasing rate and the initial condition as
long as the decoherence is present. In the study of the decoherence
effect on a system having the intermolecular interactions, the former
is observed, while the latter does not occur for all cases. The
long-time stationary state is observed to be dependent on the
decoherence strength $\Gamma$,  the initial condition of the system,
and the chiral discrimination energy parameter $v$.  Details are as
follows.
In particular, we examine three different classes of initial
conditions: the LS, WDS, and SDS cases. We apply
Eqs.(\ref{nonlinearEL3}) and (\ref{nonlinearER3}).
Coherence losses are monitored by measuring the purity rather than
$|\rho_{LR}|^2$ since the purity enables us to compare  easily the
cases with different maximum coherences and is
representation-independent. We  point out that all decoherence-free
cases were presented up to $t=5$ (Figs. \ref{figg001} to
\ref{figg04951}), while the cases with the decoherence will be
presented  up to $t=20$ to reveal the longer-time behavior. We varied
$\Gamma$ from 1.6$\times 10^{-4}$ to 1.6$\times 10^{-2}$, which
means that dephasing time is very long compared with tunneling time
(tunneling dominant region).
Figures \ref{figg501}(a) and \ref{figg501}(b) show behaviors of the
population and the purity for several values of $v$ at a very small
decoherence (here chosen as $\Gamma=1.6\times 10^{-4}$) for the LS
case. None of the cases examined here reached a stationary state
within $t=20$. As expected, introducing decoherence results in decay
of the oscillation amplitudes of both the population and the purity.
Details are affected by $|v|$: for such a small decoherence ($\Gamma
= 1.6\times 10^{-4}$), the $v=0$ case is seen to make no significant
changes within a time window of $t=20$, while cases with a nonzero
$v$ tend to experience relatively significant changes, suggesting
that the intermolecular interactions make the system more vulnerable
to  decoherence.  However, the decoherence  rate is not exactly
proportional to $|v|$. A relative sensitivity to decoherence depending
on the magnitude of $v$ is more clearly seen from comparing the purity
decay for several cases with different $v$ in Fig. \ref{figg501}(b).
The order in the rate of purity decay $R$ is
$R(v=\pm 4)>R(v=\pm 3.9)>R(v=\pm 3)> R(v=\pm 5)>R(v=\pm 10)$.
From the behavior of $\rho_{RR}$ for the decoherence-free case
shown in Fig. \ref{figg001}(a), one can see that until $|v|$ becomes
large enough to cause the self-trapping ($v=\pm 5,\,\pm 10$), the
purity decay rate $R$ increases with increasing $|v|$. However, when
the self-trapping occurs, $R$ becomes smaller than other cases that
show no self-trapping, and tends to be smaller with a stronger
self-trapping, \ie, greater $v$. This trend is also observed in the
LS case with larger $\Gamma$, $\Gamma=1.6\times 10^{-3}$ in Fig.
\ref{figg501}(c) and $\Gamma=1.6\times 10^{-2}$ in Fig.
\ref{figg301p}(a).
For instance, consider the case with $v=\pm 10$, which induces a
very strong self-trapping [Fig. \ref{figg001}(a)].  The purity decay
rate for $v=\pm 10$ is almost overlapped with that for $v=0$  with
$\Gamma=1.6\times 10^{-4}$ [Fig. \ref{figg501}(b)] and
$\Gamma=1.6\times 10^{-3}$ [Fig. \ref{figg501}(c)], and even becomes
smaller than that for $v=0$  with $\Gamma=1.6\times 10^{-2}$ [Fig.
\ref{figg301p}(a)]. It suggests that for some chiral discrimination
energies, especially those inducing a strong self-trapping
(localization), the resistance to decoherence can be stronger
than for the $v=0$  case, whereas, for values of $v$ not inducing
strong self-trapping the opposite effect is seen with respect to
decoherence, namely the non-zero $v$ case is more sensitive to
decoherence than the $v=0$ case. It is noteworthy that the effect
of chiral interactions on the LS case is very similar to that of
decoherence; for the initially localized state, namely purity decay
is accelerated with increasing decoherence, but starts to be
suppressed with continued increase in decoherence beyond a certain
threshold decoherence \cite{cplhhan_wardlaw}.
Figures \ref{figg301p}(a) and \ref{figg301p}(b) show that for
$\Gamma=1.6\times 10^{-2}$ both of the purity $(t\rightarrow
\infty)$ and $\rho_{RR}(t\rightarrow \infty)$ for the LS case tend
to be larger for a larger $|v|$, implying that for a bigger chiral
discrimination energy the system relaxes to a stationary state with
greater chirality. Specifically,  within $t=20$,  for $v=\pm 3$, the
system will be relaxed to a racemic mixture ($\rho_{RR}\rightarrow
0.5$), whereas, for $v=\pm 10$, the system will preserve its
chirality as $\rho_{RR}> 0.95$ at $t = 20$ [Fig. \ref{figg301p}(b)].
Also it is notable that the purity for $v=\pm 3$ has a stationary
value of roughly 0.57, whereas the purity for $v=\pm 10$ does not
decay much from the initial value (=1), being  0.96 at $t=20$ [Fig.
\ref{figg301p}(a)]. This can be understood by realizing that, from
comparison with the decoherence-free cases shown in Fig.
\ref{figg001}(a), $v=\pm 10$ stabilizes the initially localized state
by self-trapping it to the initially located well, and this effect
will not be much affected by the decoherence inhibiting the
oscillations between the two wells. On the other hand $v=\pm 3$
does not bring about that kind of self-trapping effect, and rather
makes the system anharmonically oscillate between the two wells,
which would be prohibited by the loss of coherences in the presence of
decoherence. For the LS case, the stronger the self-trapping, the less
vulnerability to decoherence. Also, in relation to stationary values
of the purity and $\rho_{RR}$, for $v=\pm 3$, the observation that
the purity $\rightarrow$ 0.57 and $\rho_{RR}\rightarrow 0.5$,
implies that, from Eq. (\ref{purity_LR}), $\rho_{LR}$ does not relax
to zero ($|\rho_{LR}|^2$ goes to roughly 0.04), and the system can
have therefore gained coherence due to the  combined influence of the
decoherence, the intermolecular interactions, and the initial
condition. It is also notable that decoherence brings the purity and
population ($\rho_{RR}$) for $v=\pm 3.9$ and $\pm4$ cases quite close
together, as shown in Fig. \ref{figg301p}(a) and [Fig. \ref{figg001}(b)],
respectively, smoothing out the dramatic differences between them in the
absence of decoherence [Fig. \ref{figg001}(a)].
To see the effect of increasing decoherence on the population
relaxation and the purity decay for a given $v$, we present
$\rho_{RR}$ and the purity for $v=5$, with several values of
$\Gamma$, in Fig. \ref{figv501}. As decoherence $\Gamma$ increases,
the rate of relaxations in both the population and the purity
increases. Also it is notable that both of $\rho_{RR}(t\rightarrow
\infty)$ and the purity $(t\rightarrow \infty)$ tend to decrease with
increasing $\Gamma$, implying that with increasing $\Gamma$ the $v=5$
case tends to approach a less chiral stationary state. Therefore, by
inducing the loss of coherences between the two states and thus leading
to the suppression of the oscillation (tunneling), decoherence can
bring to a system a stabilization of the chirality with $\Gamma =
1.6\times 10^{-3}$, or a less chiral state (or possibly racemization)
with $\Gamma = 1.6\times 10^{-2}$, as determined by the interplay of the
decoherence ($\Gamma$) and the intermolecular interactions ($v$).
Now let us turn to the opposite case, the initially strongly
delocalized state (SDS) case. The results of the SDS case are
presented in Fig. \ref{figg44951} with a variation of $v$ at
$\Gamma=1.6\times 10^{-3}$.   First, in regard to the relative order
of the purity decay, one can easily see that the stronger the chiral
discrimination energy ($v$), the faster the purity decays, \ie, the
larger the decoherence rate is [Fig. \ref{figg44951}]: with positive
$v$, $R(v=2)<R(v=3)<R(v=5)<R(v=10)$ [Fig. \ref{figg44951}(b)], and with
negative $v$, $R(v=-2)<R(v=-3)<R(v=-5)<R(v=-10)$
[Fig. \ref{figg44951}(d)]. Unlike the LS case, an exception is not
found in the SDS case due to non-existence of self-trapping state in the
SDS case for parameter ranges considered here [Fig. \ref{figg04951}].
Second, the SDS case with positive $v$ is seen to be more vulnerable to
decoherence than that with negative $v$. For the SDS case, as shown in
Fig. \ref{figg04951}, time-evolution of population of the initially
strongly delocalized state ($a_L\approx a_R$) depends strongly on the
sign of $v$. When there is no decoherence effect, for negative $v$, the
$\rho_{RR}$ oscillates in a narrow range about $\rho_{RR}\approx 0.5$
whereas for positive $v$, $\rho_{RR}$ oscillates over a much wider range
between $\rho_{RR}\approx 0.5 $ and $\rho_{RR}\approx 1$. That is, the
initially strongly delocalized state becomes stable for negative $v$
but unstable for positive $v$. This feature continues to be effective
upon introducing the decoherence effect, and thus the SDS case shows
a faster purity decay and a faster population relaxation for
positive $v$ than for negative $v$
[Fig. \ref{figg44951}]. It suggests that a critical difference in
the stability of the initial states, depending on the sign of $v$,
\ie, whether the heterochiral interactions are stronger or weaker
than the homochiral interactions, affects the sensitivity of the
system to the decoherence effects. This is reconfirmed in the WDS
case ($a_L\ll a_R$) for which results are presented in Fig.
\ref{figg40595} with a variation of $v$ at $\Gamma = 1.6\times
10^{-3}$.  In contrast to the SDS case [Fig. \ref{figg44951}], the
WDS case with negative $v$ is more vulnerable to decoherence than
that with positive $v$. For the WDS case in the absence of
decoherence, when the homochiral mixture is preferred (the positive
$v$), the initially weakly delocalized state becomes stabilized,
while, when the heterochiral mixture is preferred (negative $v$), the
initial state becomes highly unstable and the system leads to the
facilitated tunneling [Fig. \ref{figg00595}]. Thus, upon
introducing the decoherence,  the system with positive $v$ shows a
less vulnerability to the decoherence, \ie, a slower purity decay
and a slower population relaxation, than the corresponding cases
with  negative $v$ [Fig. \ref{figg40595}]. It is very interesting to
see that the character of intermolecular interactions can change the
sensitivity of the system to the decoherence, by (de)stabilizing the
initial state.  On the other hand, for  the WDS case,  the relative
sensitivity to the decoherence seems to have a rather complicated
dependence on $v$ due to its medium sized initial coherence,
compared with the two extreme cases, the LS and SDS cases.
The effects of increasing decoherence on the WDS and SDS cases for a
given $v$ are demonstrated in Figs. \ref{figv50595} and
\ref{figv54951}, respectively. Similarly to the LS case, for a
given $v$, in both of the WDS and SDS cases, one can see that the
larger the decoherence is, the faster the purity decays and the
faster the population relaxes. Further, related to the
homochiral/heterochiral preferences in the intermolecular
interactions, introducing decoherence does not necessarily lead to
the breakdown of the homochiral/heterochiral preference. For the WDS
case, for instance, a value of $v=5$ yields a system with a
homochiral preference when $\Gamma=1.6\times 10^{-3}$
($\rho_{RR}(t\rightarrow \infty)$ is expected to be greater than 0.9
as shown in Fig. \ref{figv50595}(a)), while a value of $v=-5$ results
in a racemic mixture without chirality, reflecting its heterochiral
preference for the same value of $\Gamma$ [Fig. \ref{figv50595}(c)].
Similarly, for the SDS case, at all $\Gamma$ examined here, for $v=5$
$\rho_{RR}\rightarrow 0.51$ [Fig. \ref{figv54951}(a)], implying a small
homochiral preference. For $v=-5$ $\rho_{RR}$ is expected to relax to
0.5 with oscillations around 0.5, implying the heterochiral preferences
[Fig. \ref{figv54951}(c)].
\section{conclusion}\label{conclusion}
We studied the effects of chiral intermolecular
interactions and decoherence on the distinguishability between the
initially pure and mixed states, and the chirality stabilization in
a chiral system.  A two-level approximation was adopted for the
system.  Chiral intermolecular interactions were introduced within a
mean-field theory.  The interaction of the system with an
environment (to take  account for externally induced decoherence)
was modeled as a continuous position measurement of the system,
which reduces to a measurement of a population difference between
the two chiral states. To better understand the combined effects of
the two interactions, we first investigated the system in the
presence of each interaction separately.
For the system with decoherence only, most features is studied in
Ref.\cite{cplhhan_wardlaw} where we obtained the following results
(which we presented in summarized form in this work). When
$Y_0^{\mrm{P}}$ is nonzero, for the dephasing dominant region,
increasing the tunneling frequency can lead to long-time survival of
distinguishability, while for the tunneling dominant region, the
timescale of the distinguishability is largely dominated by the
dephasing rate. For the initially $\k R$ state (LS), decreasing the
tunneling frequency tends to suppress the purity decay, suggesting
that the more classical  system is less vulnerable to decoherence.
In the system with chiral intermolecular interactions only, we
showed that the nonzero chiral discrimination energies can bring
about the distinguishability between the initially pure and the
corresponding mixed states even when $Y_0^{\mrm{P}}=0$, where no
distinguishability exists in the absence of the intermolecular
interactions.
Finally, we introduced both decoherence and intermolecular
interactions to the system, and then explored the combined effects
on the purity decay rate and the stationary states of the system
with several different initial conditions. The dephasing time was
varied to be relatively very long compared with tunneling time
(tunneling dominant region) so that the decoherence process in the
absence of chiral intermolecular interactions is very slow. However,
when combined with the chiral intermolecular interactions, the
decoherence process tends to be accelerated. For a given chiral
discrimination energy $v$, we showed that, similar to the case
with decoherence only, the purity decays faster as the dephasing
rate increases. For a given dephasing rate, the purity decay
was shown to be influenced by $v$ and the initial state of the
system. As the chiral discrimination energy $|v|$ increases, the SDS
case is observed to have a faster purity decay, while the LS case to
have a purity decay suppressed beyond a certain threshold $|v|$
because of onset of  induced self-trapping. This is very similar to
the interplay of the initial coherence of the system and the
dephasing rate on the purity decay in the system with decoherence
only.  Also we clarified the role of the sign of $v$ associated with
the initial state on the purity decay.  For the WDS case, the purity
is seen to decay slower with positive $v$ than with negative $v$
since positive $v$, which prefers homochiral mixtures, stabilizes
the initially weakly delocalized state.  For the SDS case, the
opposite is true.
The results should serve as a prototype for understanding the
results of a chiral system in more complicated and realistic
conditions, such as systems with more than two degrees of freedom
(beyond a two-level approximation) and with other types of
environment, for instance, which induces both population loss
(energy relaxation) and phase loss (dephasing).
This work may be extended to an open chiral system interacting with
lasers.  In recent years, there has been increasing interest on the
control of molecular chirality
\cite{harris_cina_romero,control_brumer_control_ohtsuki}. However,
the decoherence effect and the intermolecular interactions  on these
preparation/control scenarios are far from well understood.
Realizing that the decoherence process can be detrimental to the
control scenarios by destroying the very coherence that is needed
for the control scenario, it would be critical to understand and
include the decoherence process properly when the time scales
of system dynamics, and/or environmental effects (energy loss,
dephasing) are comparable to the interaction time scale between the
system and the laser field. In addition,  our results suggest that
taking into account these effects may be important even when the
decoherence time scale is not comparable to or longer than other
time scales, since a system with nonzero chiral discrimination
energies tends to be more vulnerable to decoherence, \ie, tends to
have a faster purity decay than a system without chiral intermolecular
interactions.
\begin{center}
    {\bf Acknowledgements}
\end{center}
It is a pleasure to acknowledge the Natural Sciences and Engineering
Research Council of Canada for financial support in the form of a Discovery
Grant.

\begin{figure}[htbp]
\centerline{\hbox{\epsfxsize=4.5in \epsfbox{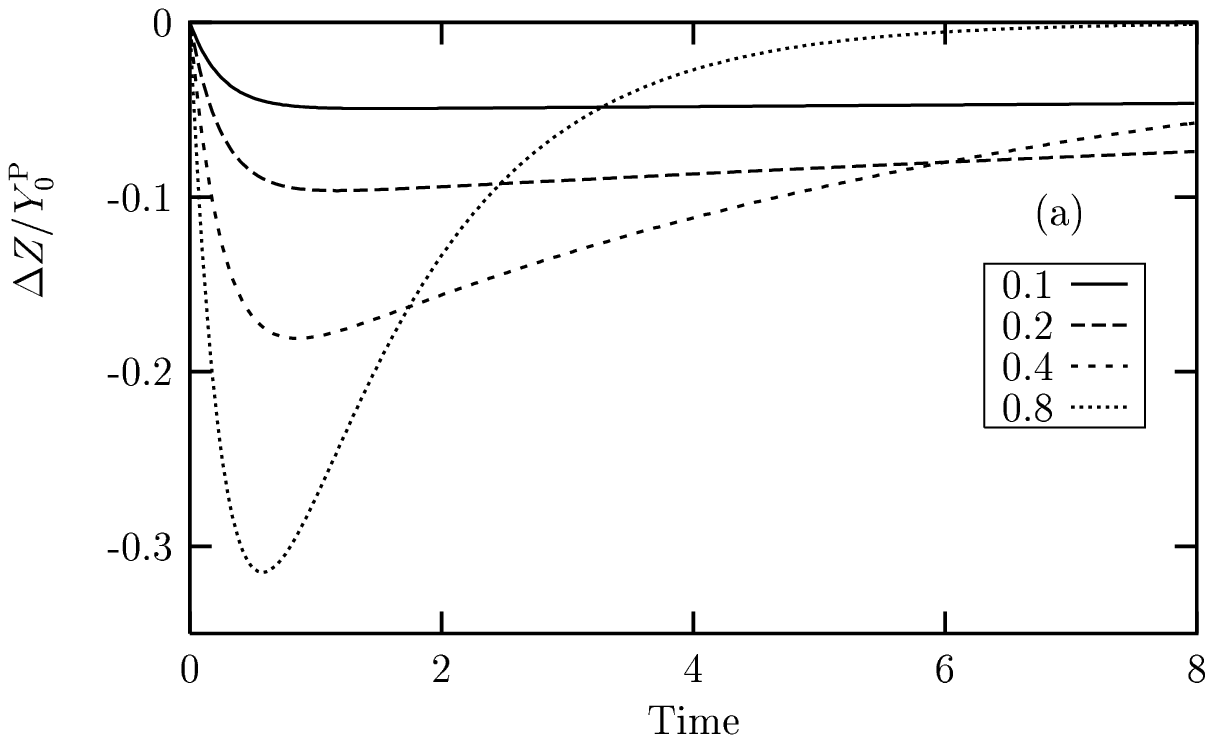}}}
\centerline{\hbox{\epsfxsize=4.5in \epsfbox{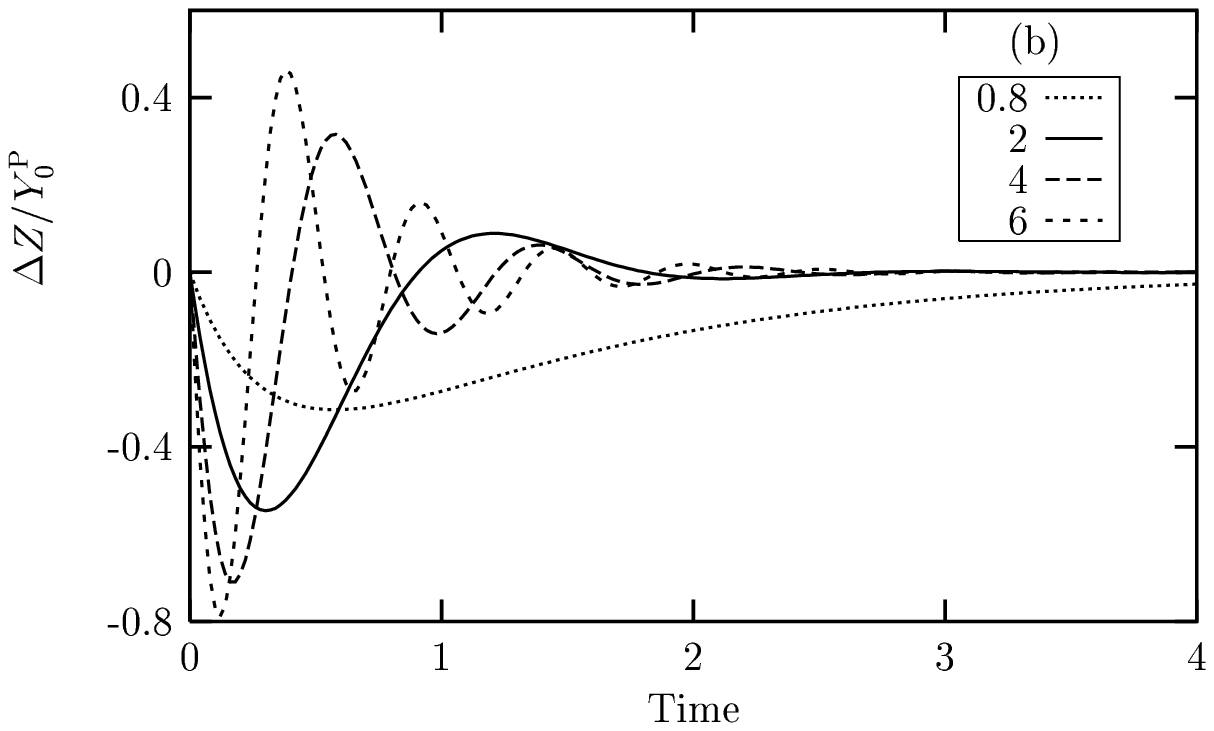}}}
\centerline{\hbox{\epsfxsize=4.5in \epsfbox{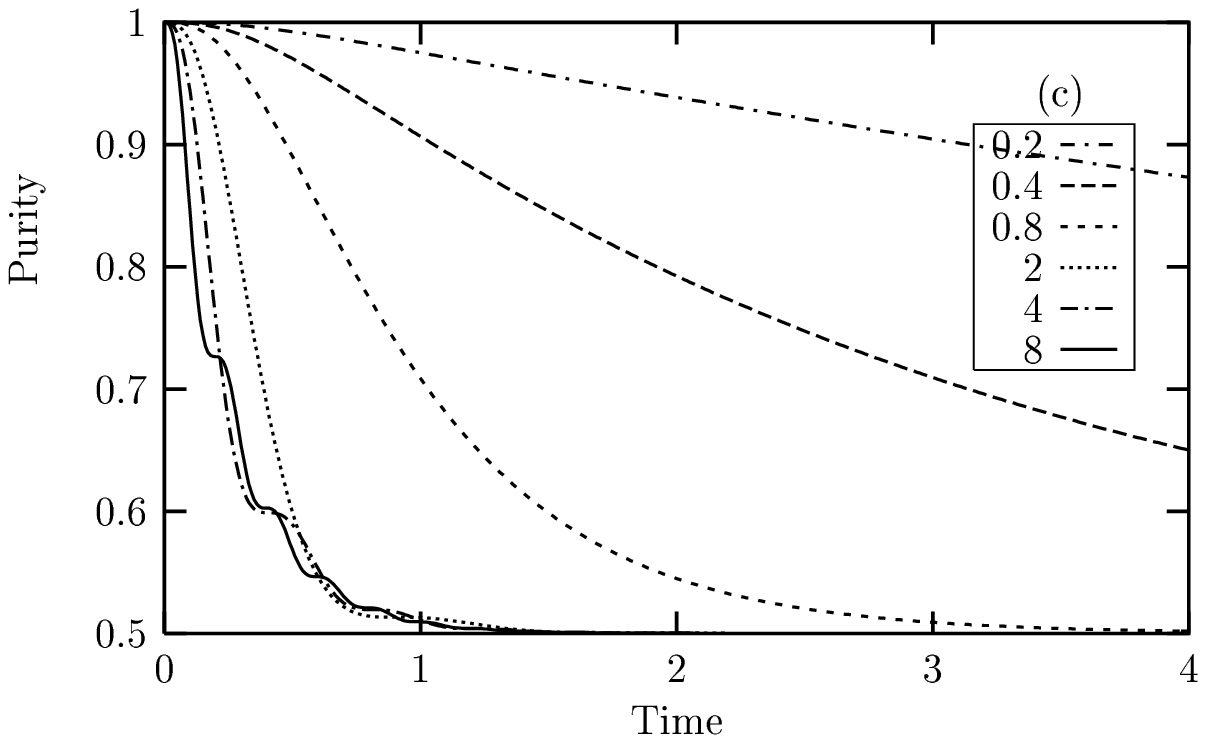}}}
\caption{}
\label{newddeltaz}
 \end{figure}

\newpage
\centerline{{\bf FIGURE  CAPTIONS}}
\vspace{0.2in} \noindent Fig. \ref{newddeltaz}. Dependence of
distinguishability and decoherence process on tunneling frequency
$\omega$ as a function of time for the decoherence only case.
Distinguishability versus time as measured by $\Delta Z (t)(\equiv
Z^{\mrm{P}}(t)- Z^{\mrm{M}}(t))/ Y_0^{\mrm{P}}$ is depicted for the
dephasing dominant region $\omega < 1$ in panel (a) and for
tunneling dominant region $\omega > 1$ in panel (b). The values of
$\omega$ are indicated in the legend and included in both panels to
provide a comparison. Purity vs. time is depicted in panel (c) for a
range of $\omega$ values; $\gamma$ is chosen as 1. All the variables
are in dimensionless units.

\begin{figure}[htbp]
\centerline{\hbox{\epsfxsize=6.0in \epsfbox{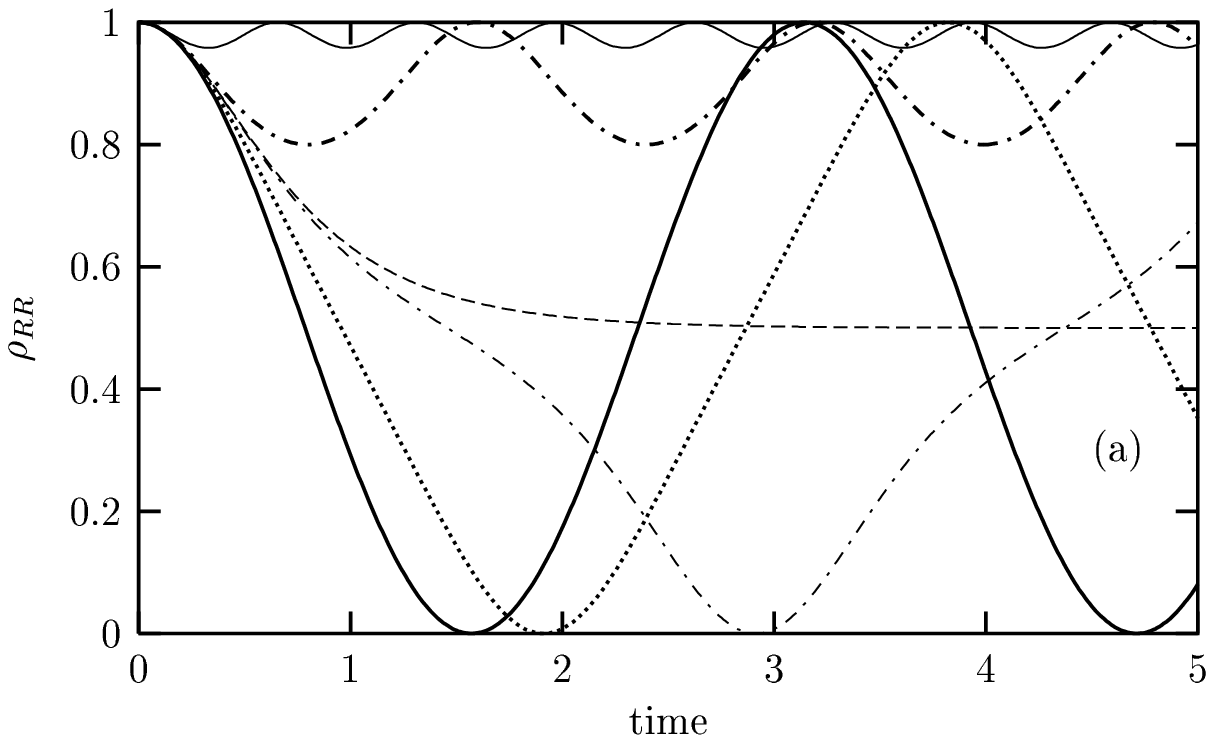}}}
\centerline{\hbox{\epsfxsize=6.0in \epsfbox{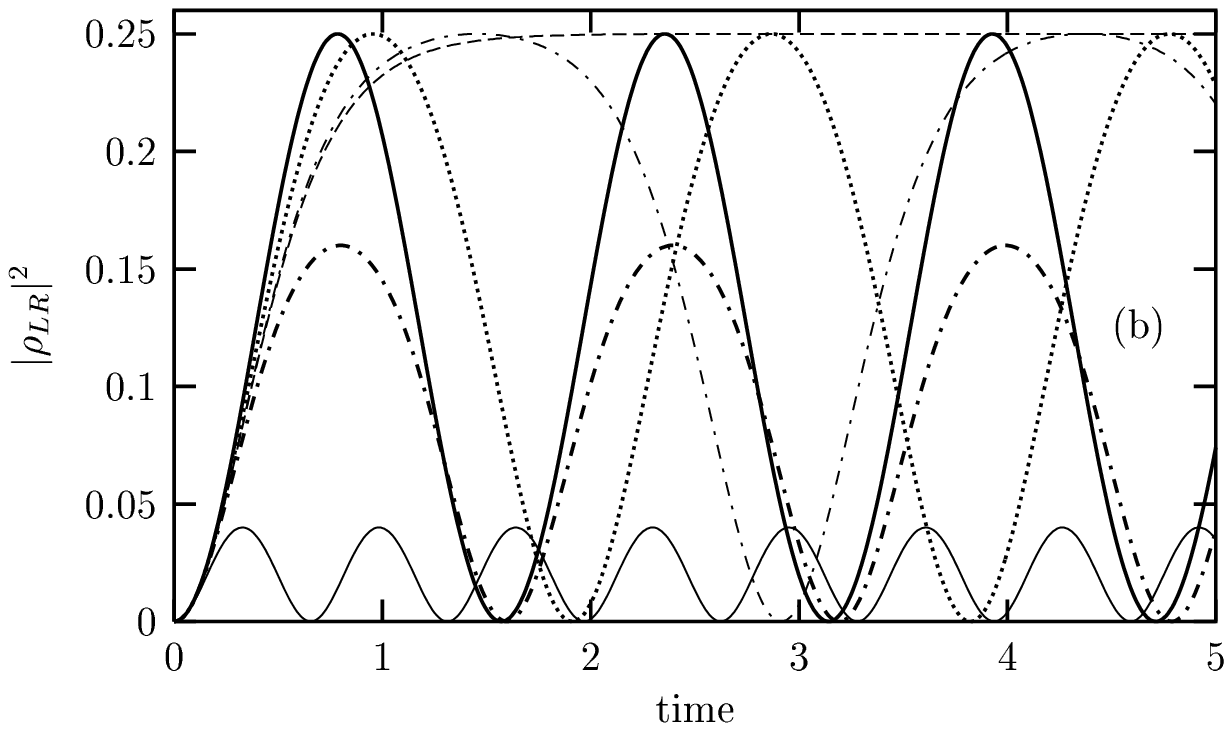}}}
\caption{}
 \label{figg001}
 \end{figure}

\vspace{0.2in} \noindent Fig. \ref{figg001}. Time evolution of population
$\rho_{RR}$ and coherence  $|\rho_{LR}|^2$ are shown for the LS case with
$v=0$ (thick line), $\pm$3 ($\cdot\cdot\cdot$), $\pm $3.9 ($-\cdot-$),
$\pm$4($---$), $\pm$5({\bf - . - . -}), and  $\pm$10 (thin line).

\begin{figure}[htbp]
\centerline{\hbox{\epsfxsize=6.0in \epsfbox{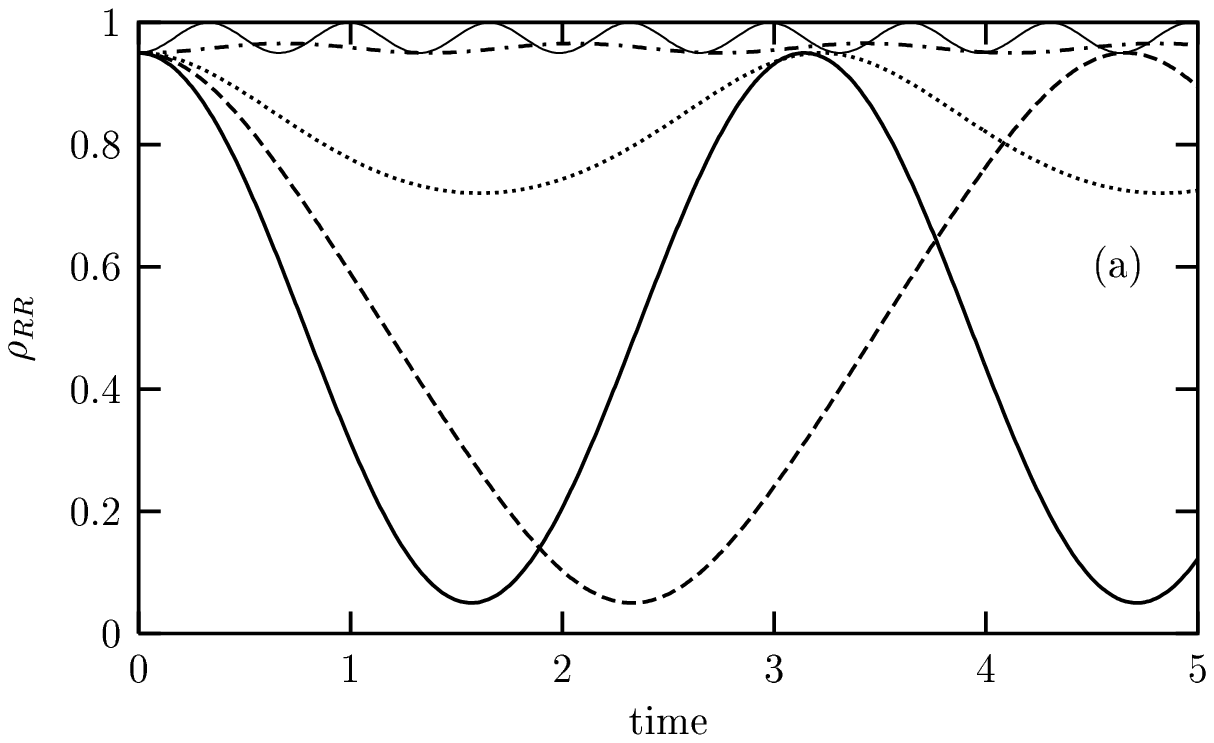}}}
\centerline{\hbox{\epsfxsize=6.0in \epsfbox{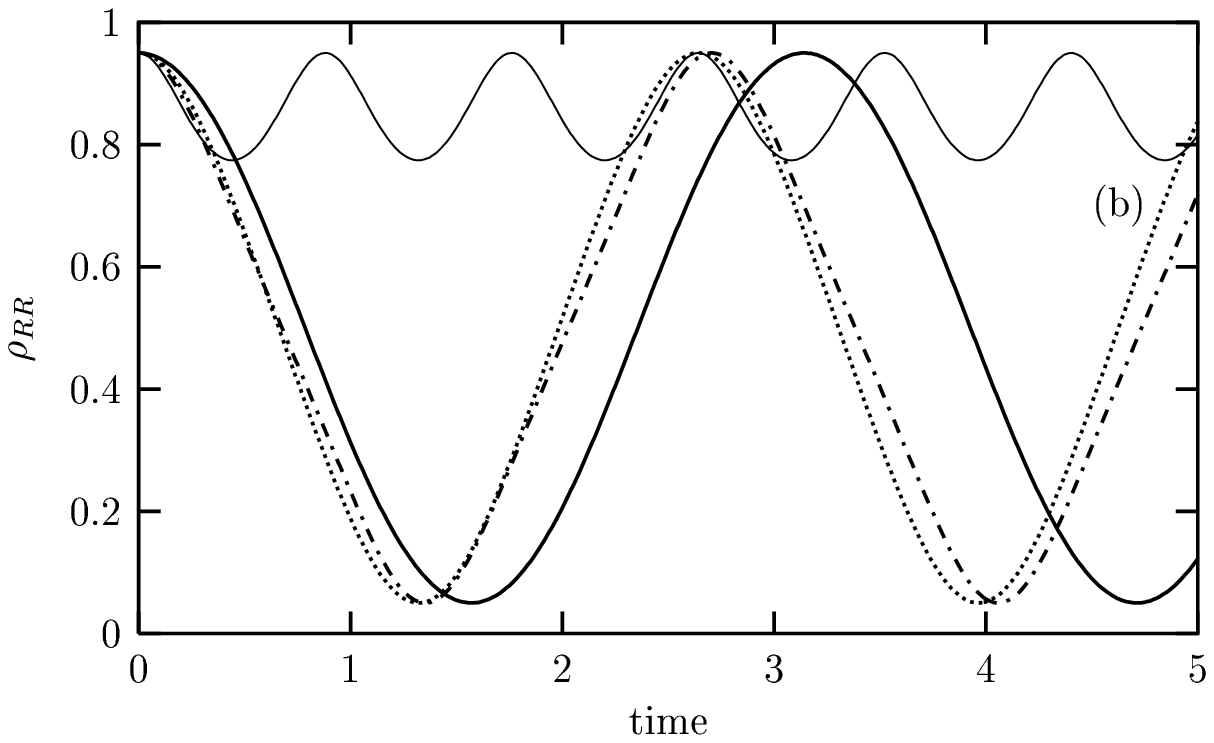}}}
\caption{}
 \label{figg00595}
 \end{figure}

\vspace{0.2in} \noindent Fig. \ref{figg00595}. Time evolution of
$\rho_{RR}$ is shown for the WDS case with various values of $v$.
(a): $v=0$ (thick line), 2 ({\bf - - - -}), 3 ($\cdot\cdot\cdot$),
5({\bf - . - . -}), and 10 (thin line). (b): $v=0$ (thick line), -2
({\bf - - - -}), -3 ($\cdot\cdot\cdot$), -5({\bf - . - . -}), and
-10 (thin line).

\begin{figure}[htbp]
\centerline{\hbox{\epsfxsize=6.0in \epsfbox{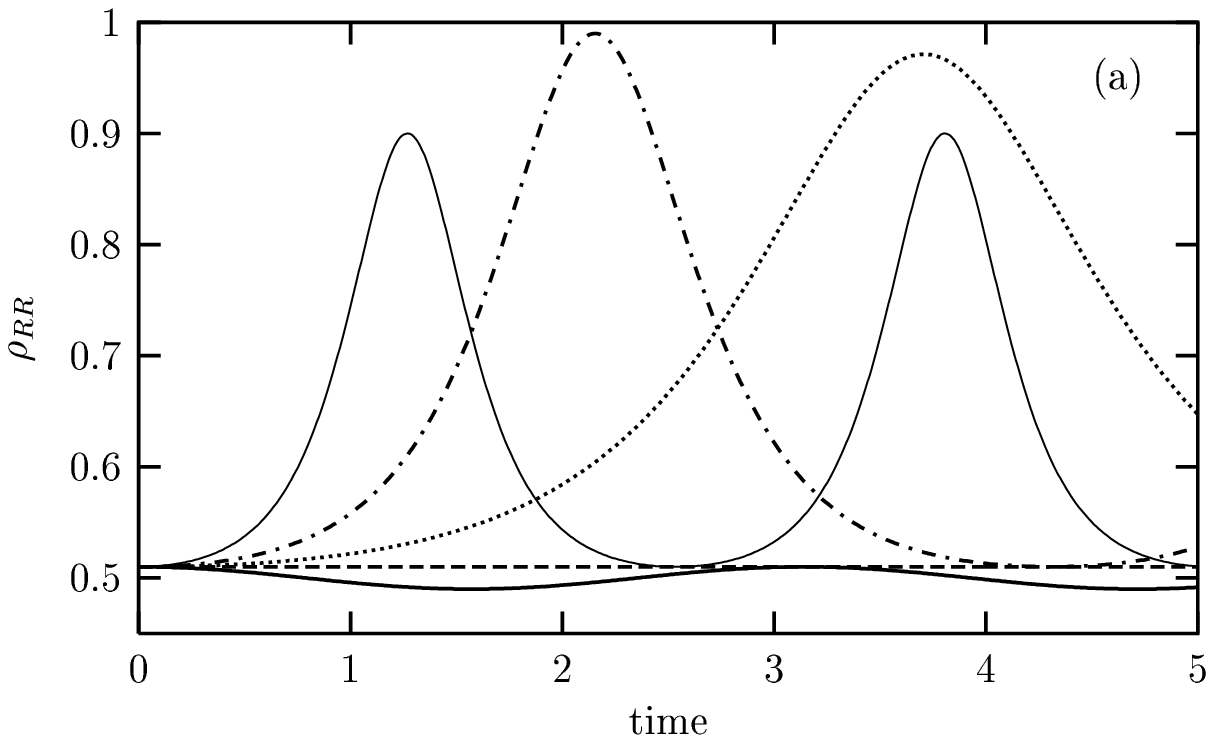}}}
\centerline{\hbox{\epsfxsize=6.0in \epsfbox{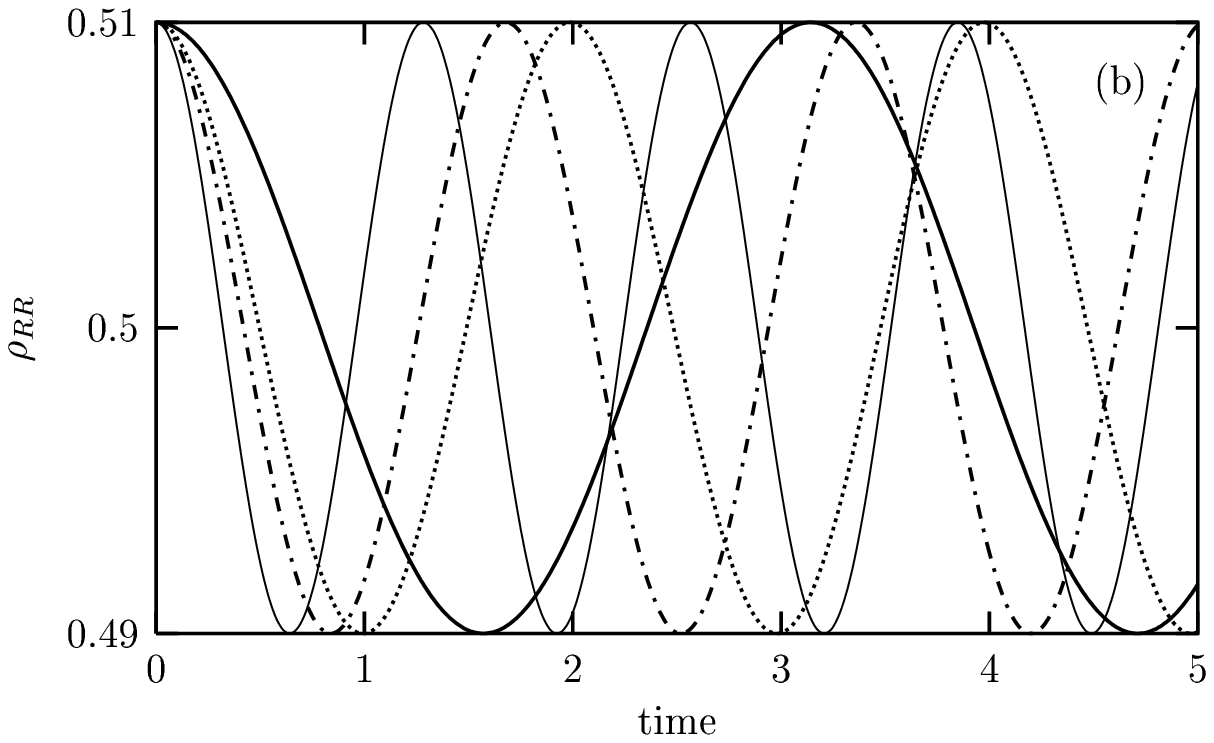}}}
\caption{}
 \label{figg04951}
 \end{figure}

\vspace{0.2in} \noindent Fig. \ref{figg04951}. Time evolution of
$\rho_{RR}$ is shown for the SDS case with various values of $v$.
(a): $v=0$ (thick line), 2 ({\bf - - - -}), 3 ($\cdot\cdot\cdot$),
5({\bf - . - . -}), and 10 (thin line). (b): $v=0$ (thick line),
-10 (thin line). Note the difference in the scale of ordinate in
panels (a) and (b).

\begin{figure}
\centering
\begin{tabular}{cc}
\epsfig{file=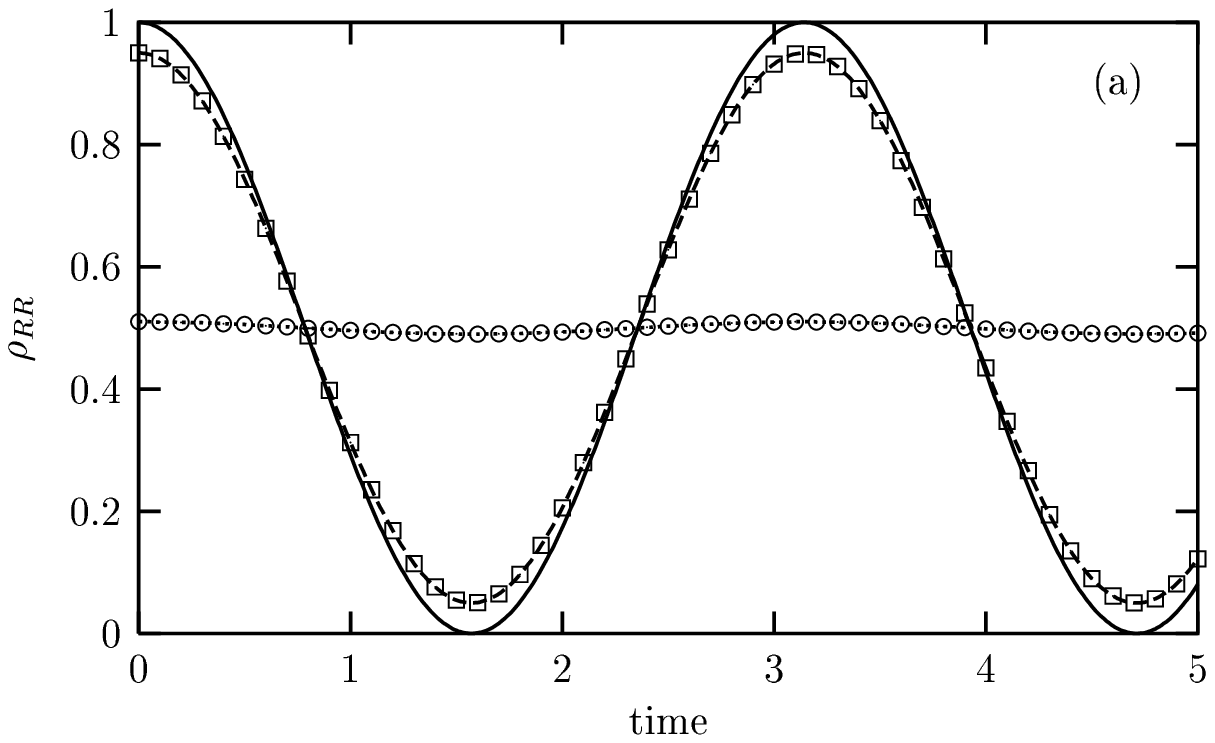,width=0.5\linewidth,clip=} &
\epsfig{file=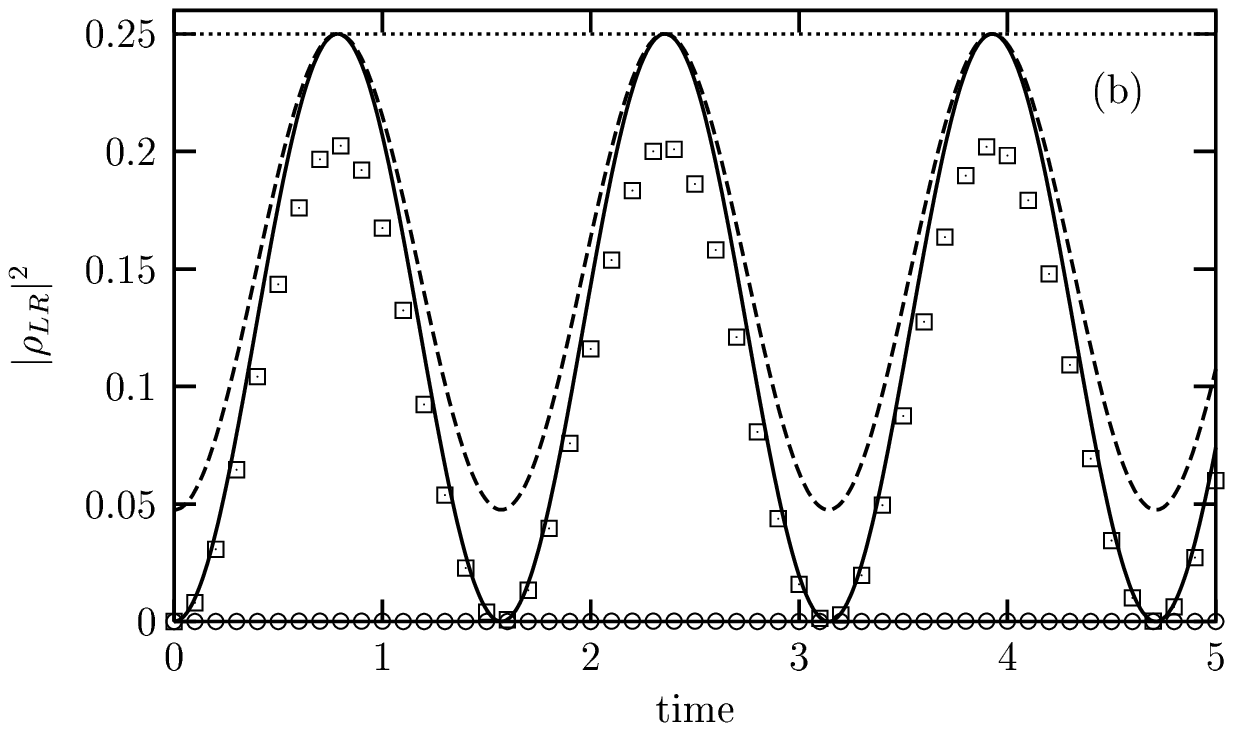,width=0.5\linewidth,clip=} \\
\epsfig{file=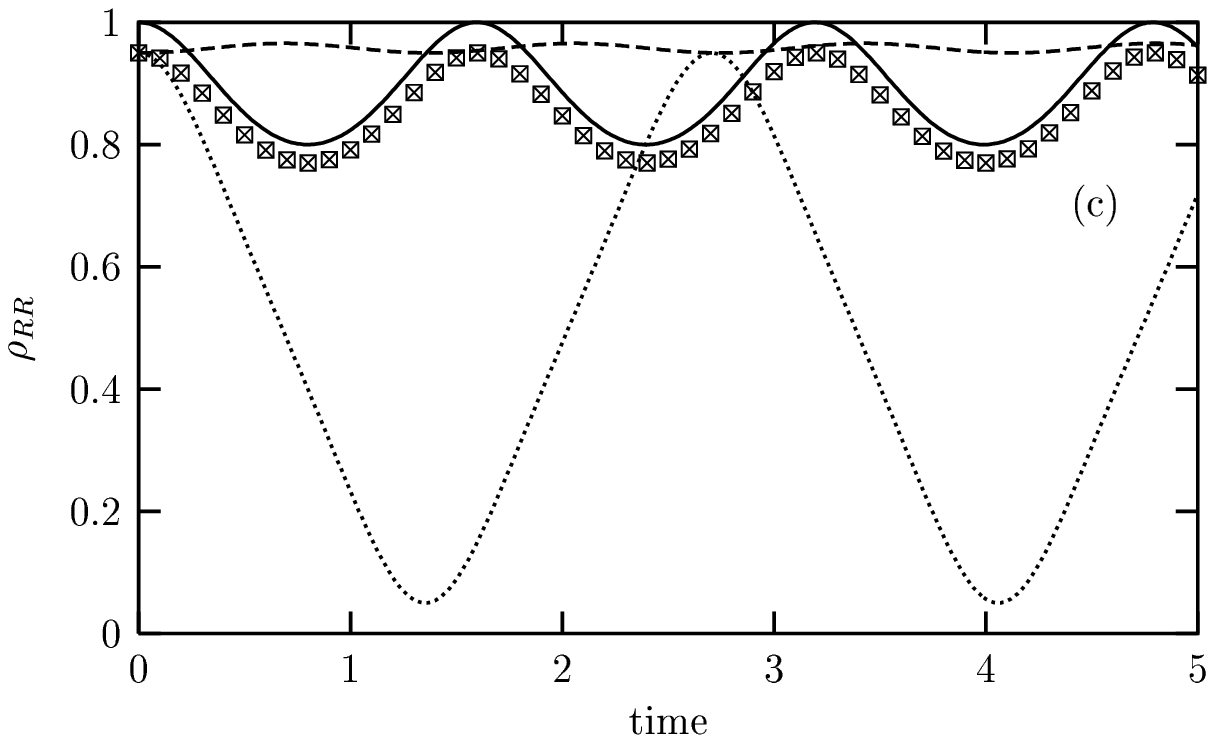,width=0.5\linewidth,clip=} &
\epsfig{file=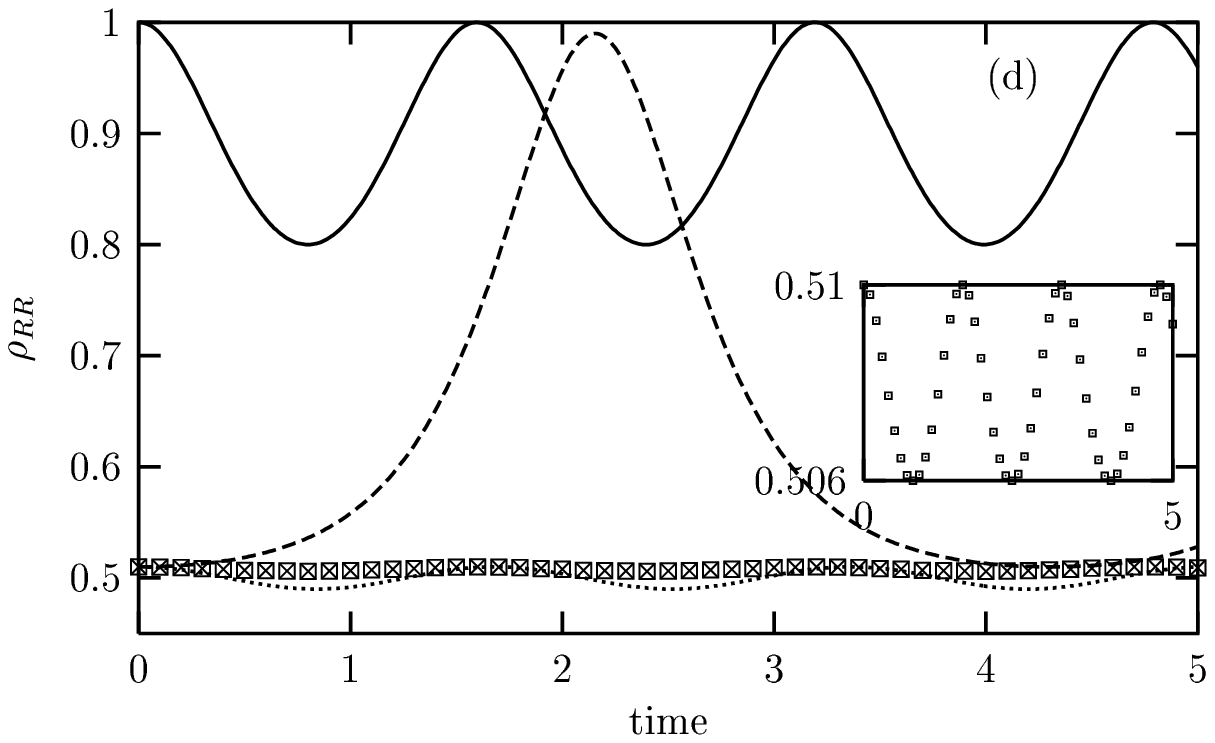,width=0.5\linewidth,clip=}
\end{tabular}
\caption{}
 \label{figg0v0mp}
\end{figure}

\vspace{0.2in} \noindent Fig. \ref{figg0v0mp}. Time evolution of
$\rho_{RR}$ and $|\rho_{LR}|^2$ are shown with various initial
conditions.  For (a) and (b), $v=0$ and four different initial
conditions are presented: the initially pure state for the WDS
case $(---)$, the initially pure state $(\cdot\cdot\cdot)$ for
the SDS case, and the corresponding mixed states are presented as
(squares) and (circles), respectively. For comparison, the LS case
with $v=0$ is also presented ({\bf ------}). Panels (c) and (d)
display the WDS and SDS cases, respectively, with four different
conditions: the initially pure states with $v=5$ $(---)$ and $v=-5$
$(\cdot\cdot\cdot)$  and the corresponding mixed states as (squares) and
(xxx), respectively. For comparison, the LS case with $v=5$, is
also presented ({\bf ------}). The inset in (d) is for the SDS case
with the initially mixed state for $v=5$ (squares) with enlarged
ordinate which reveals that it oscillates in tune with the LS case
with $v=5$.

\begin{figure}[htbp]
\centerline{\hbox{\epsfxsize=4.0in \epsfbox{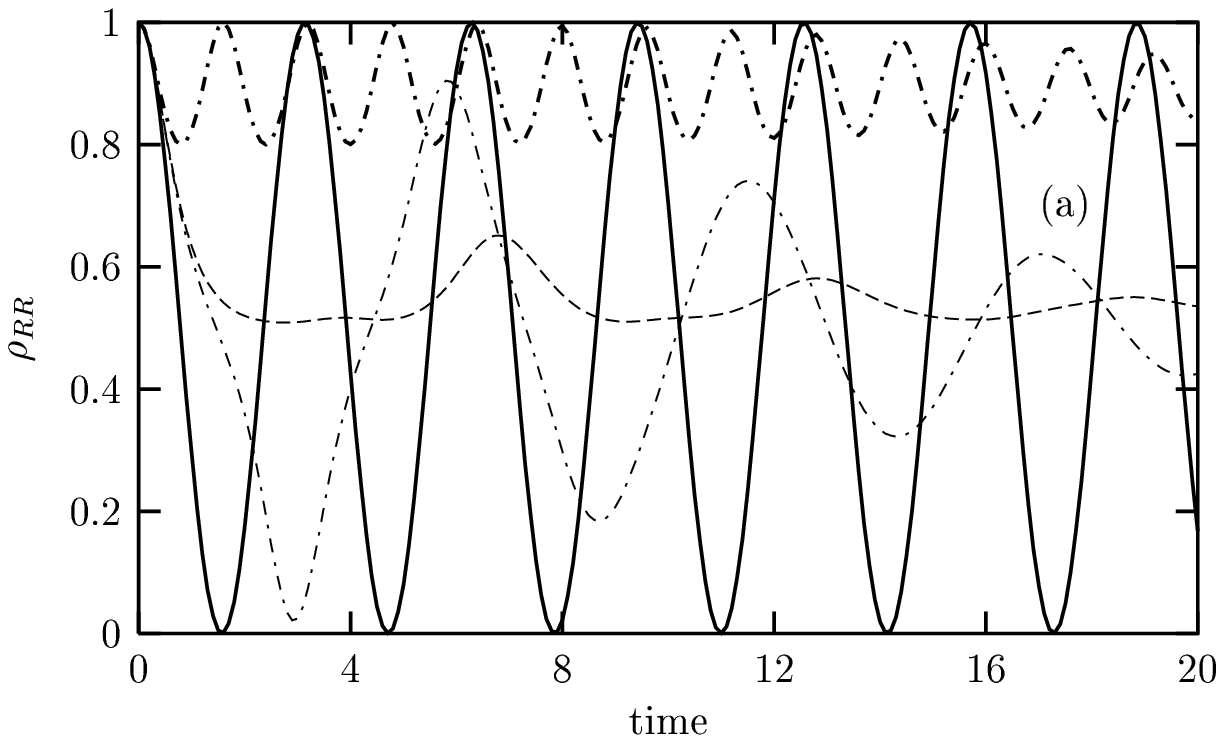}}}
\centerline{\hbox{\epsfxsize=4.0in \epsfbox{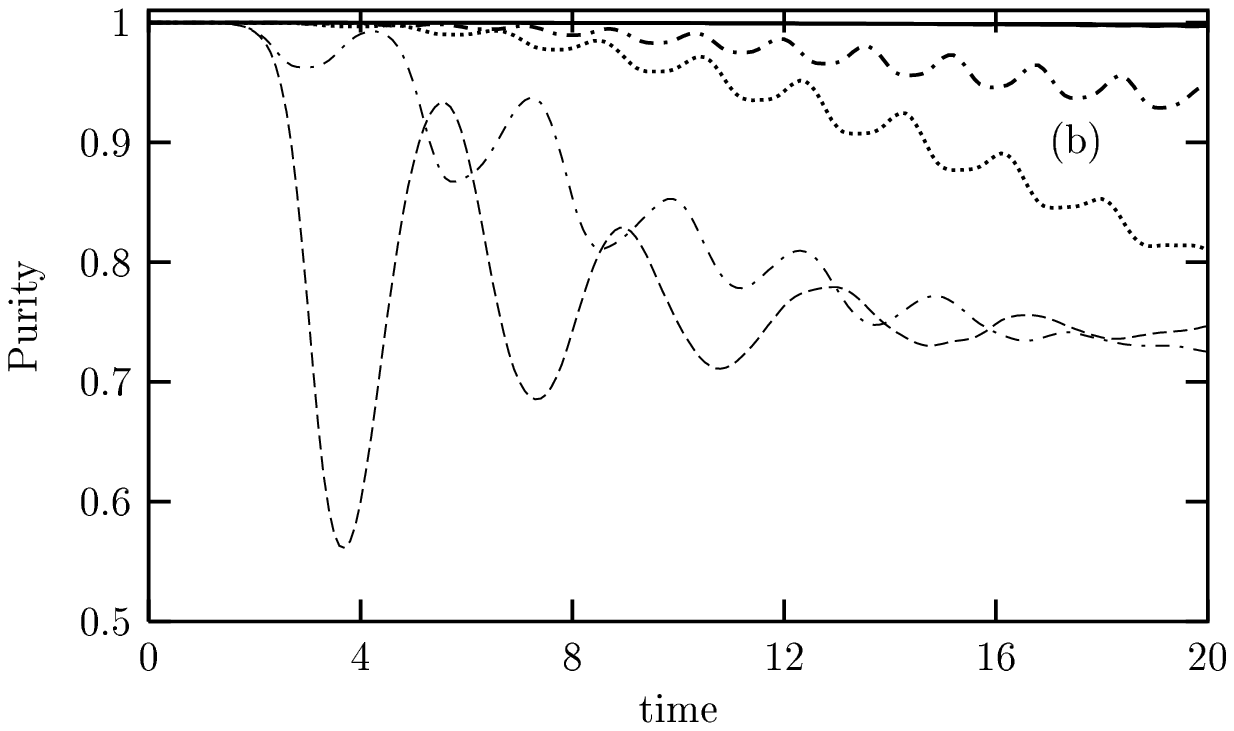}}}
\centerline{\hbox{\epsfxsize=4.0in \epsfbox{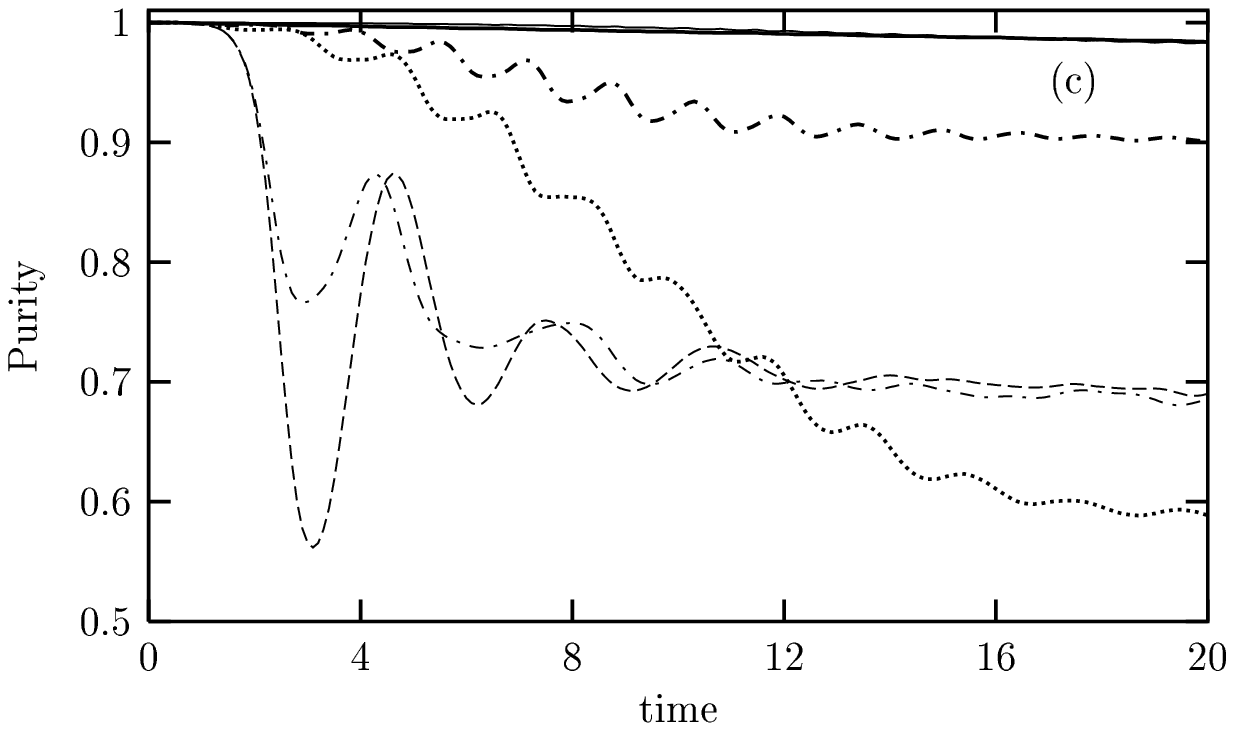}}}
\caption{}
 \label{figg501}
 \end{figure}

\vspace{0.2in} \noindent Fig. \ref{figg501}. The LS case with
various $v$ with $\Gamma=1.6\times 10^{-4}$ for (a),(b) and with
$\Gamma=1.6\times 10^{-3}$ for (c). In (a), $\rho_{RR}$ vs. the
rescaled time with $v=0$ (thick line), $\pm $3.9 ($-\cdot-$),
$\pm$4($---$), and $\pm$5({\bf - . - . -}). In (b) and (c),
purity vs. the rescaled time with $v=0$ (thick
line), 
$\pm$3 ($\cdot\cdot\cdot$),  $\pm $3.9 ($-\cdot-$), $\pm$4($---$),
$\pm$5({\bf - . - . -}), and $\pm$10 (thin line). Note that the
case with $v=\pm$10 overlapps with the $v = 0$ case.

\begin{figure}[htbp]
\centerline{\hbox{\epsfxsize=5.0in \epsfbox{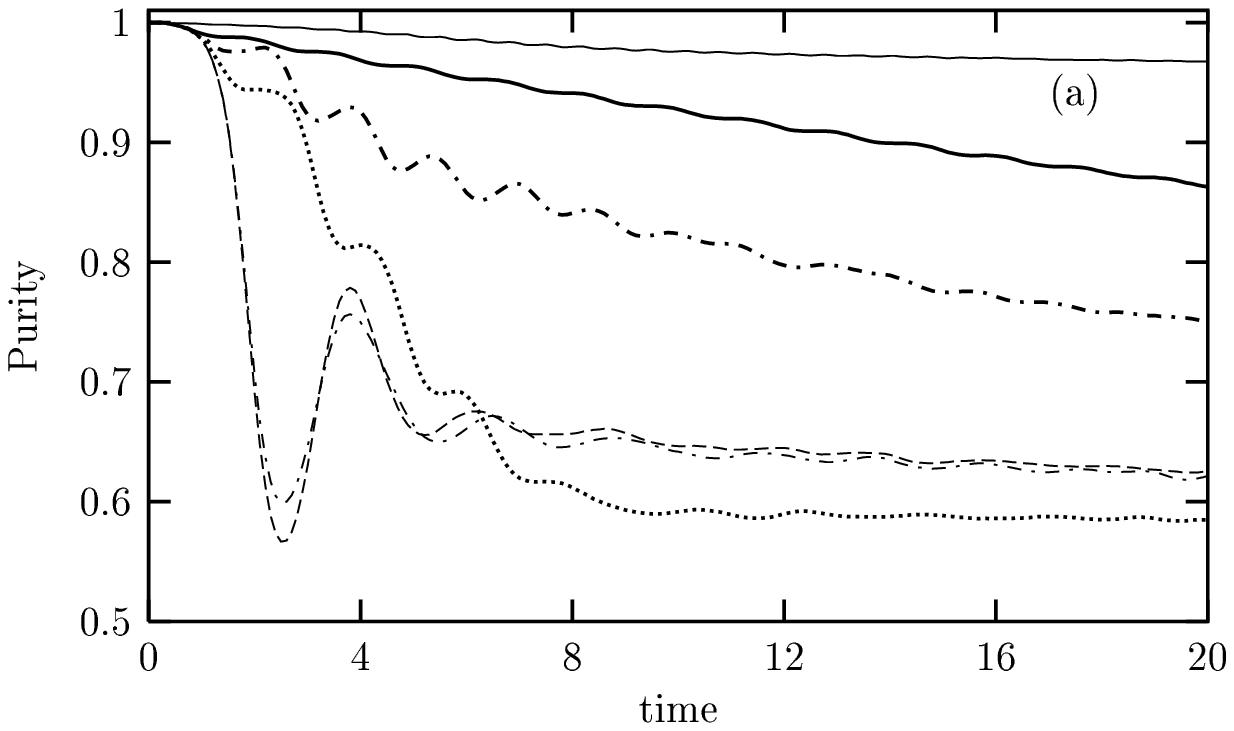}}}
\centerline{\hbox{\epsfxsize=5.0in \epsfbox{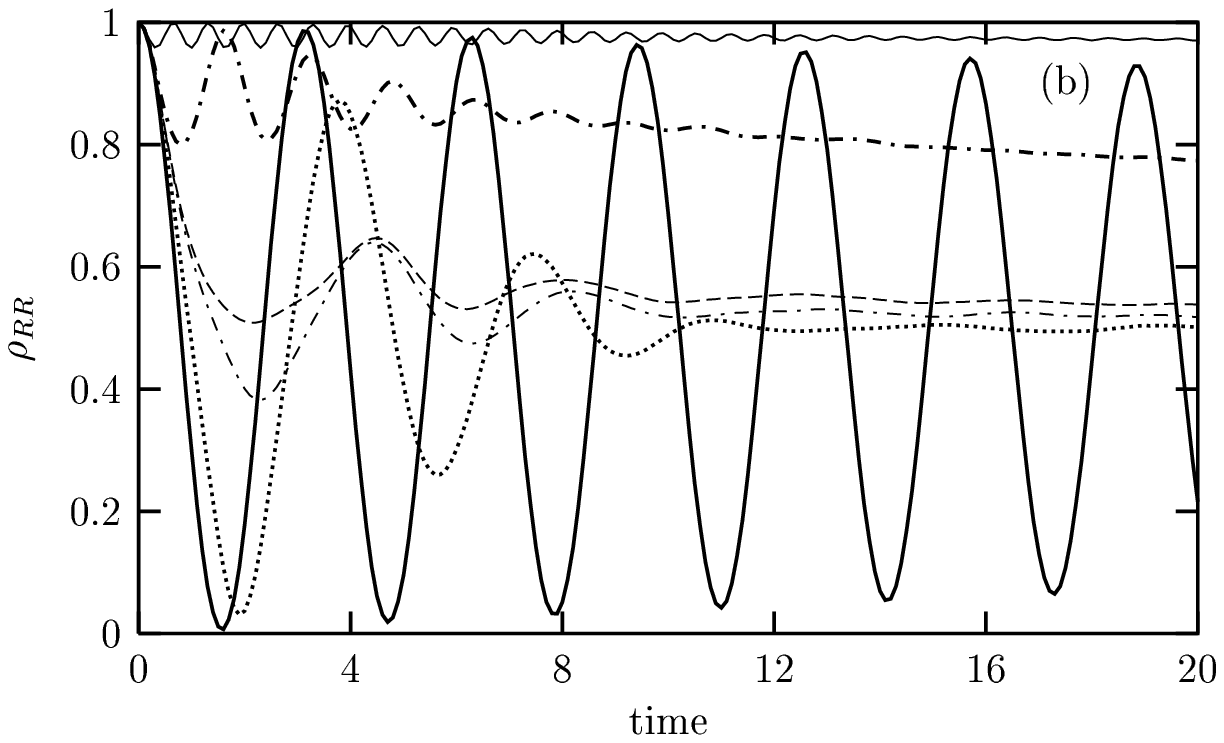}}}
\caption{}
 \label{figg301p}
 \end{figure}

\vspace{0.2in} \noindent Fig. \ref{figg301p}. Time evolution of (a)
purity and (b) population with $\Gamma=1.6\times 10^{-2}$, for the LS
case, at $v=0$ (thick line),
($\cdot\cdot\cdot$),  $\pm $3.9 ($-\cdot-$),  $\pm$4($---$),
$\pm$5({\bf - . - . -}) and  $\pm$10 (thin line).

\begin{figure}[htbp]
\centerline{\hbox{\epsfxsize=6.0in \epsfbox{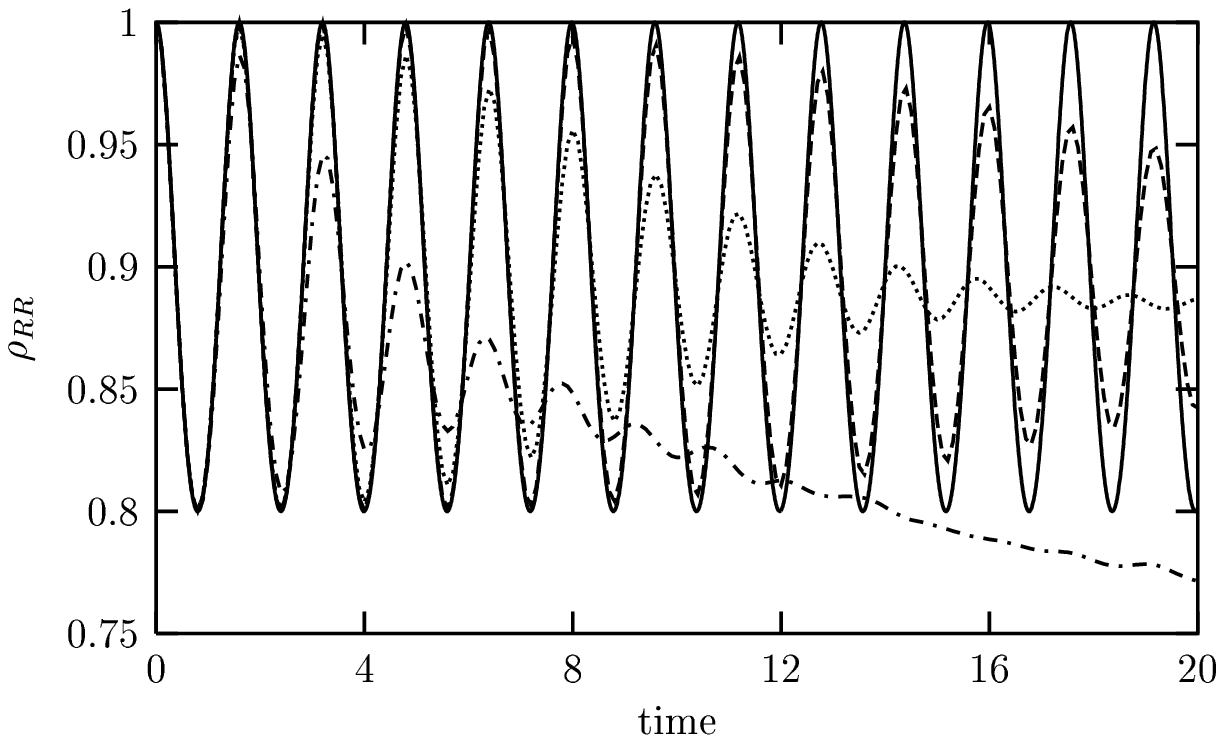}}}
\centerline{\hbox{\epsfxsize=6.0in \epsfbox{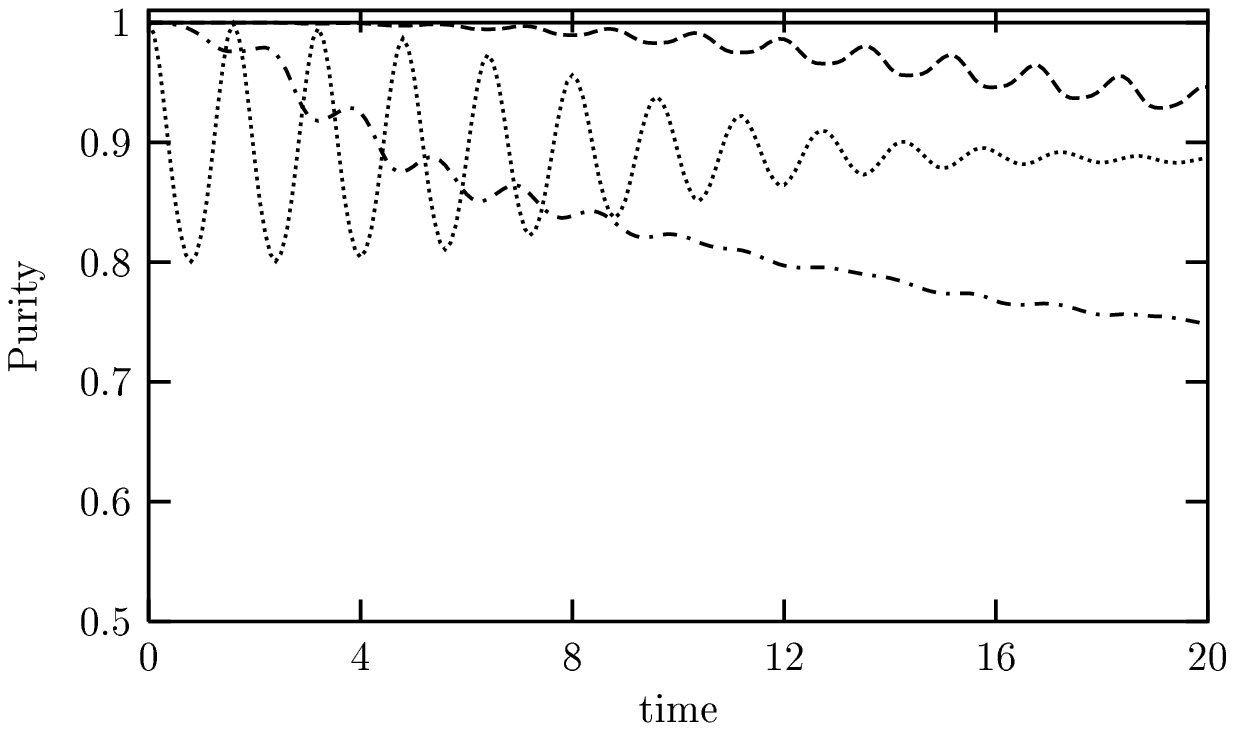}}} \caption{}
 \label{figv501}
 \end{figure}

\vspace{0.2in} \noindent Fig.  \ref{figv501}. Upper panel:
$\rho_{RR}$ vs. the rescaled time.  Lower panel: Purity vs. time.
Here the LS case with $v = + 5$ is presented at $\Gamma=0$ (thick
line), $1.6\times 10^{-4}$ ({\bf - - - -}), $1.6\times 10^{-3}$
($\cdot\cdot\cdot$), and $1.6\times 10^{-2} $ ($-\cdot-$).

\begin{figure}
\centering
\begin{tabular}{cc}
\epsfig{file=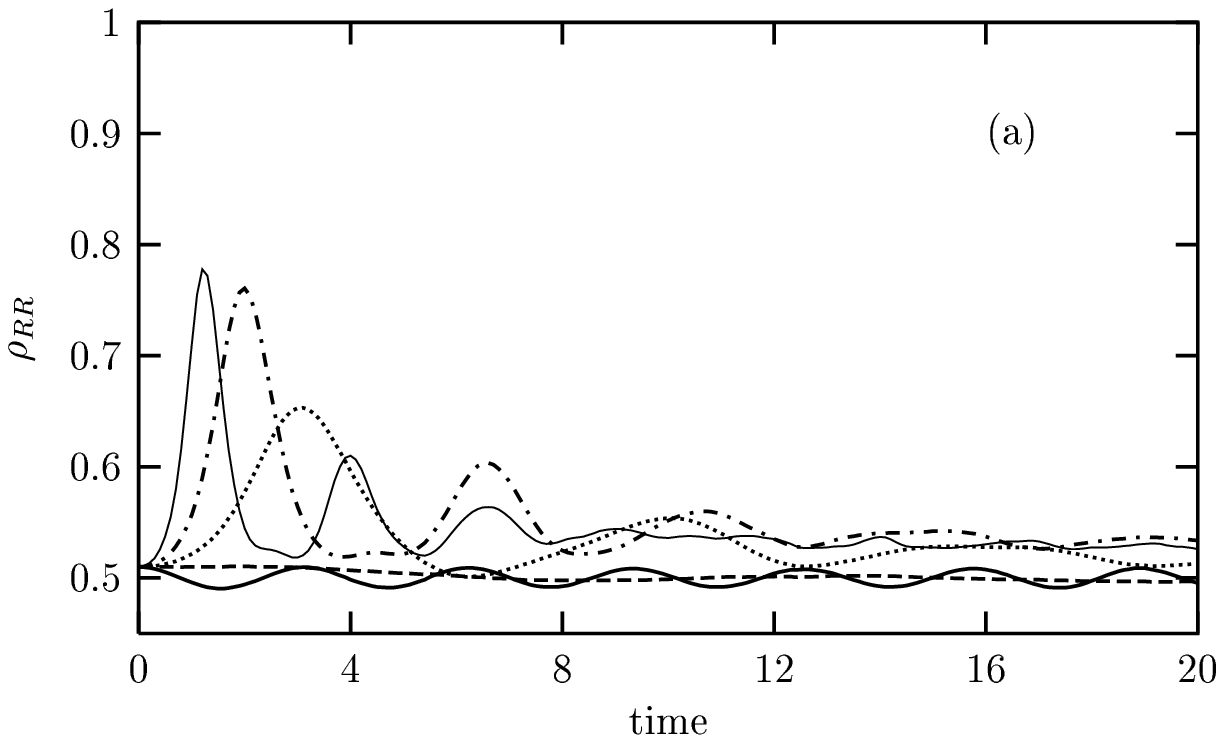,width=0.5\linewidth,clip=} &
\epsfig{file=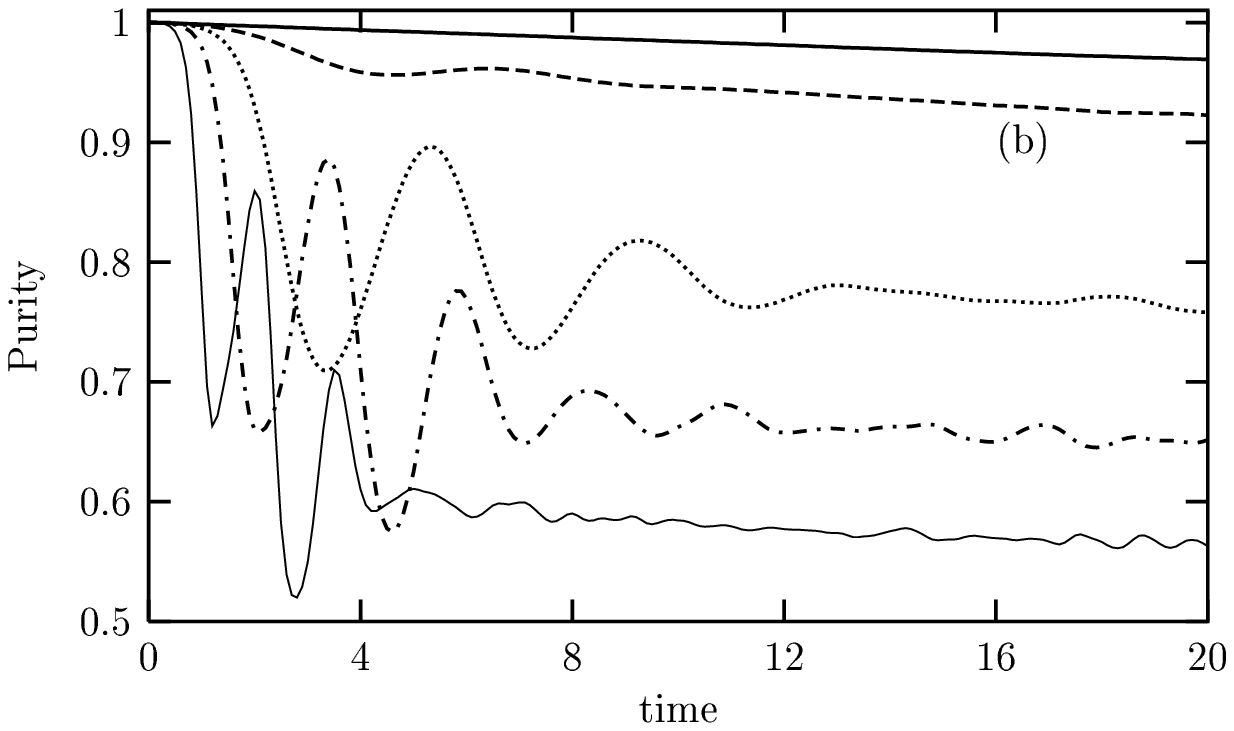,width=0.5\linewidth,clip=} \\
\epsfig{file=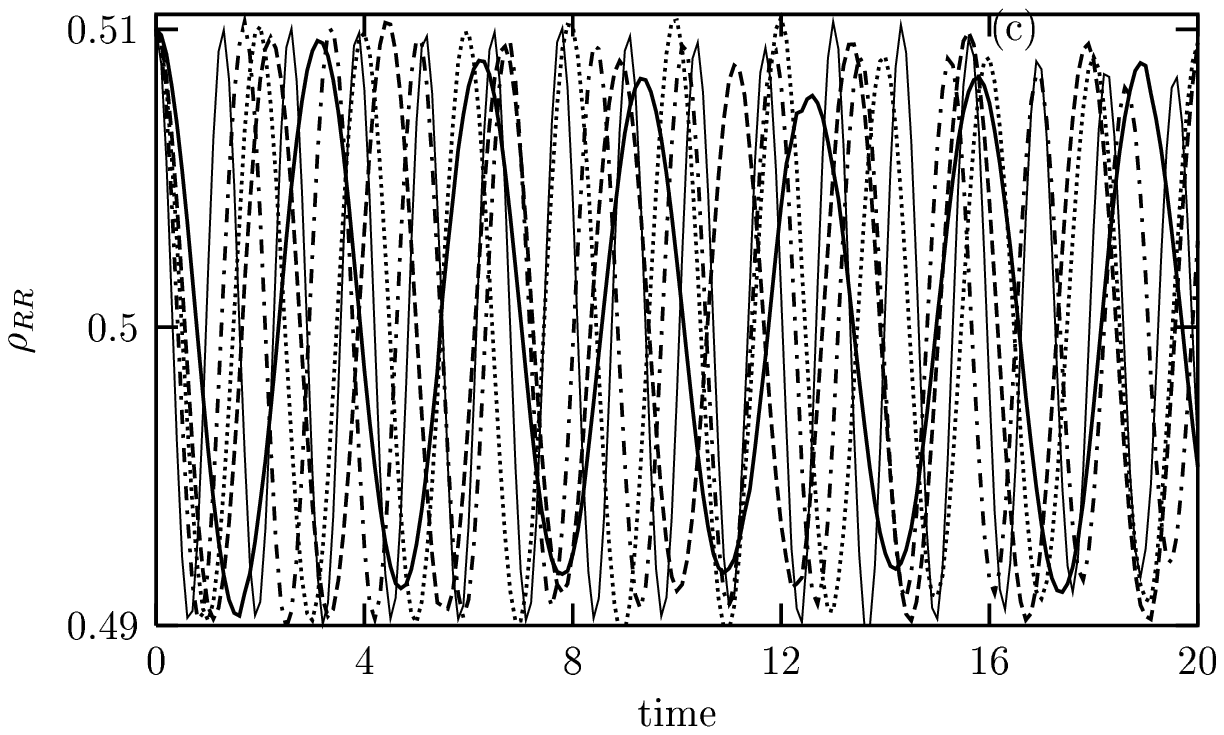,width=0.5\linewidth,clip=} &
\epsfig{file=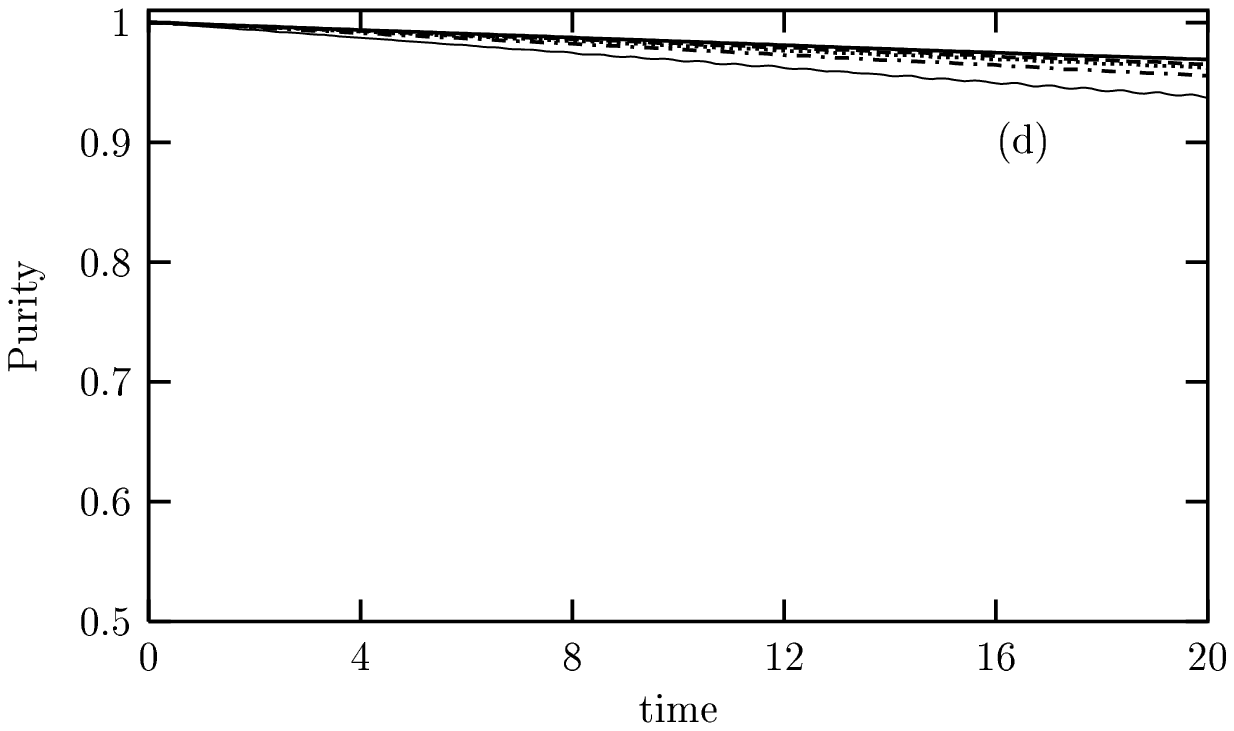,width=0.5\linewidth,clip=}
\end{tabular}
\caption{}
 \label{figg44951}
\end{figure}

\vspace{0.2in} \noindent Fig.  \ref{figg44951}. $\rho_{RR}$ vs. the
rescaled time in (a) and (c).  Purity  vs. the rescaled time in (b)
and (d).  The SDS case with $\Gamma=1.6\times 10^{-3}$ is presented.
For (a) and (b), $v>0$: $v=0$ (thick line), 2 ({\bf - - - -}), 3
($\cdot\cdot\cdot$), 5({\bf - . - . -}), and  10 (thin line). For
(c) and (d), $v<0$: $v=0$ (thick line), -2 ({\bf - - - -}), -3
($\cdot\cdot\cdot$), -5({\bf - . - . -}), and  -10 (thin line).
Note the difference in the scale of ordinate in panels (a) and (c).

\begin{figure}
\centering
\begin{tabular}{cc}
\epsfig{file=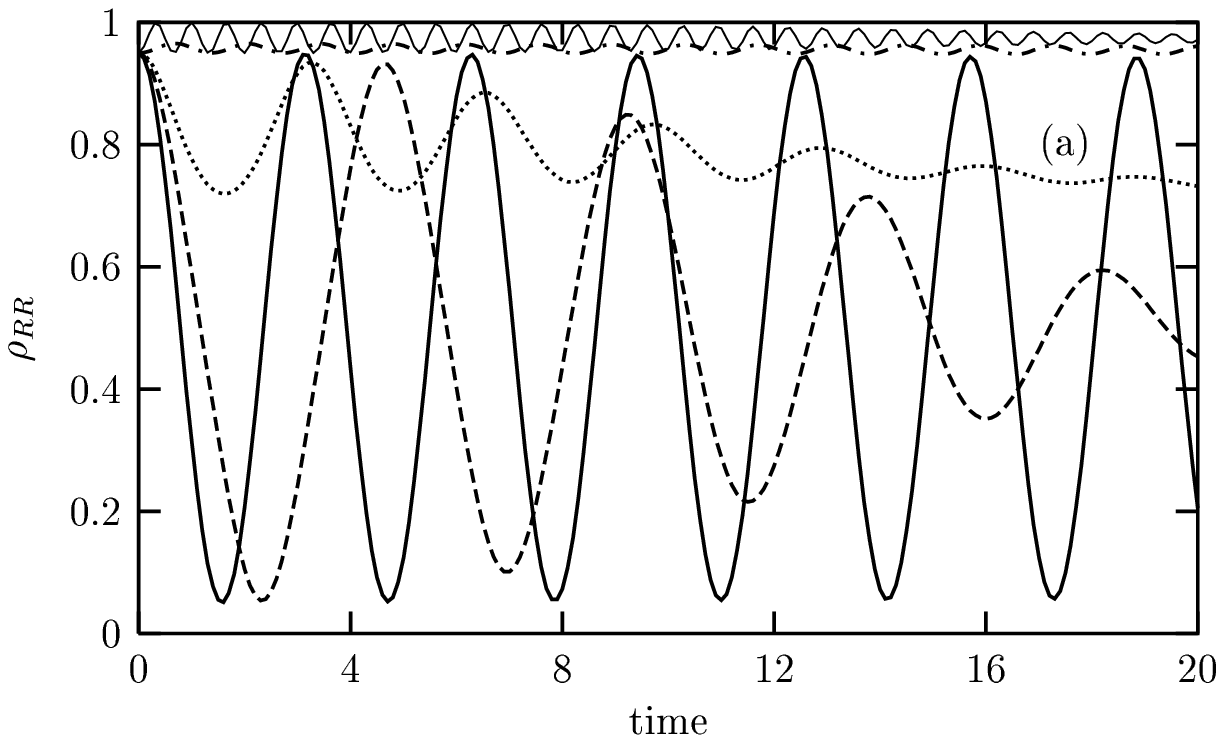,width=0.5\linewidth,clip=} &
\epsfig{file=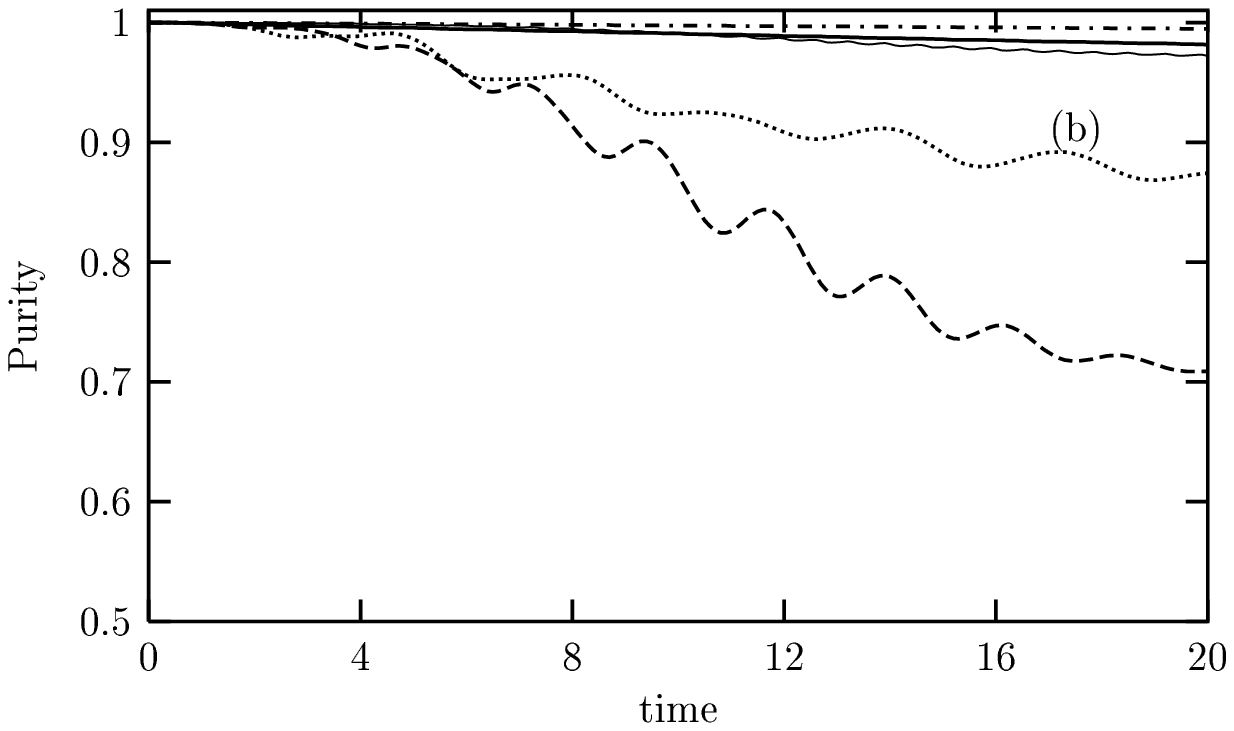,width=0.5\linewidth,clip=} \\
\epsfig{file=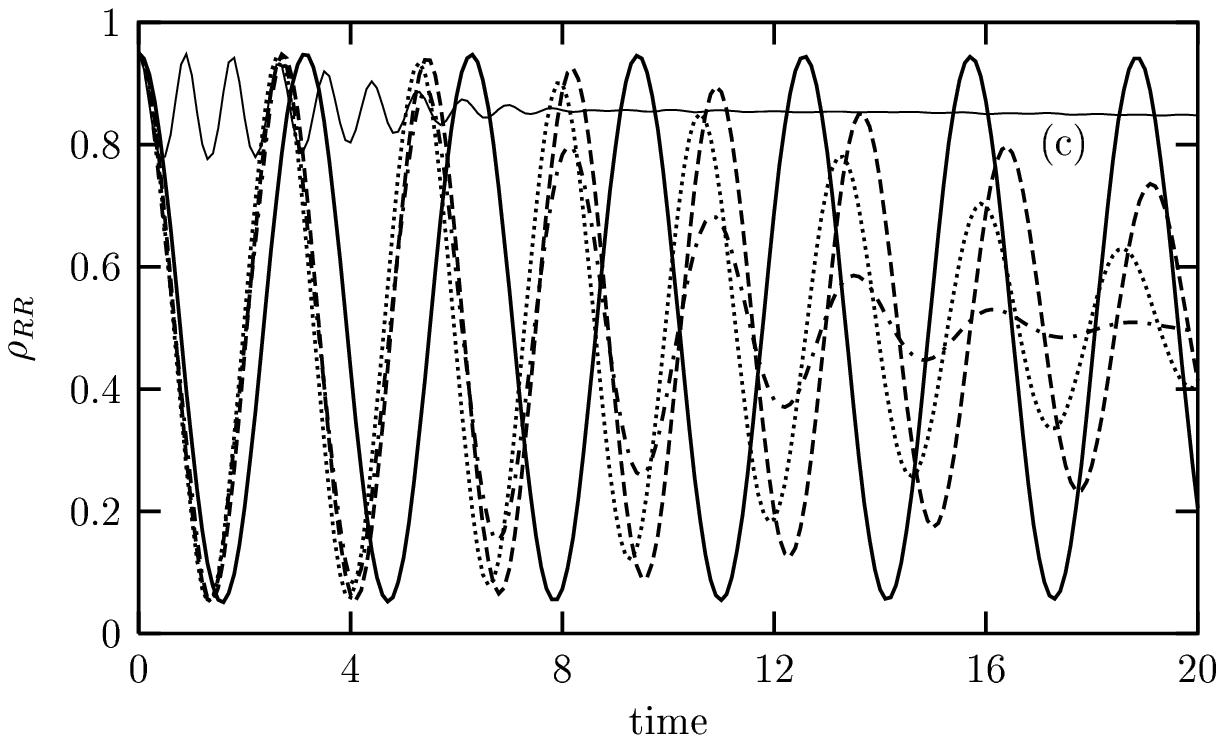,width=0.5\linewidth,clip=} &
\epsfig{file=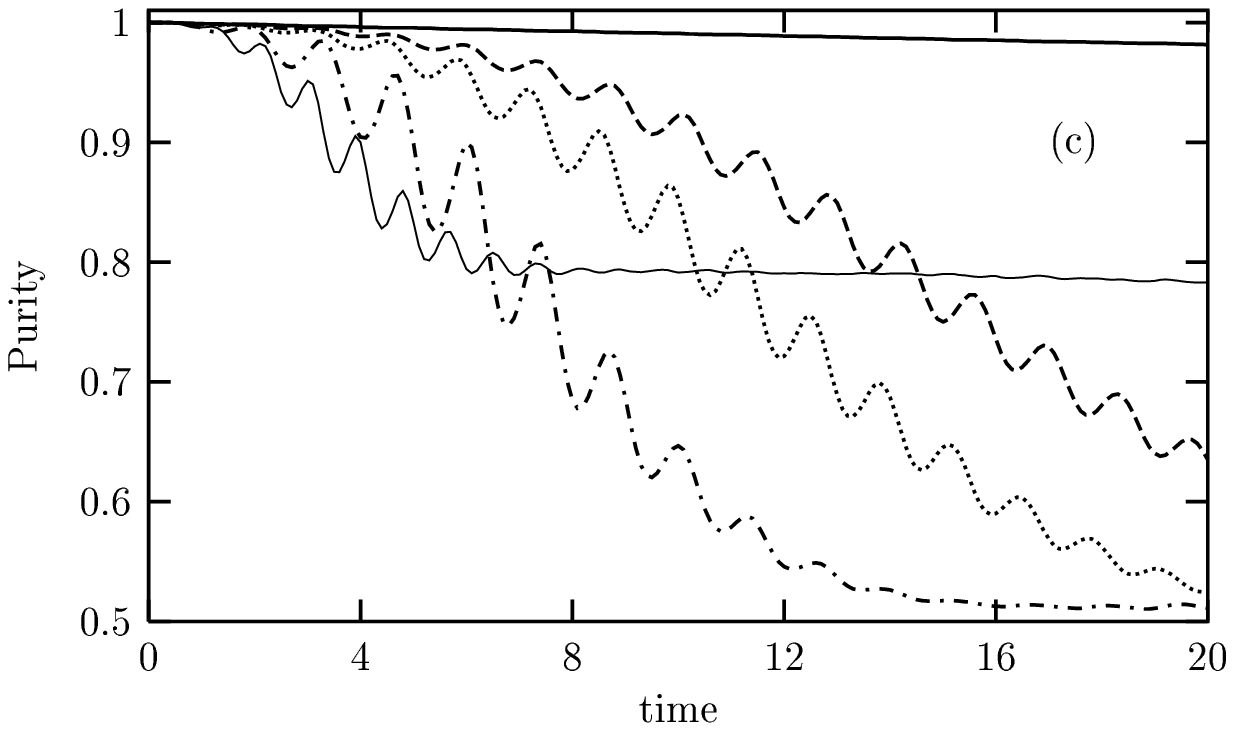,width=0.5\linewidth,clip=}
\end{tabular}
\caption{}
 \label{figg40595}
\end{figure}

\vspace{0.2in} \noindent Fig.  \ref{figg40595}.  $\rho_{RR}$ vs.
the rescaled time in (a) and (c).  Purity vs. the rescaled time in
(b) and (d).  The WDS case  with $\Gamma=1.6\times 10^{-3}$ is
presented.  For (a) and (b), $v>0$; $v=0$ (thick line), 2 ({\bf -
- - -}), 3 ($\cdot\cdot\cdot$), 5({\bf - . - . -}), and  10 (thin
line). For (c) and (d), $v<0$; $v=0$ (thick line), -2 ({\bf - - -
-}), -3 ($\cdot\cdot\cdot$), -5({\bf - . - . -}), and  -10 (thin
line).

\begin{figure}
\centering
\begin{tabular}{cc}
\epsfig{file=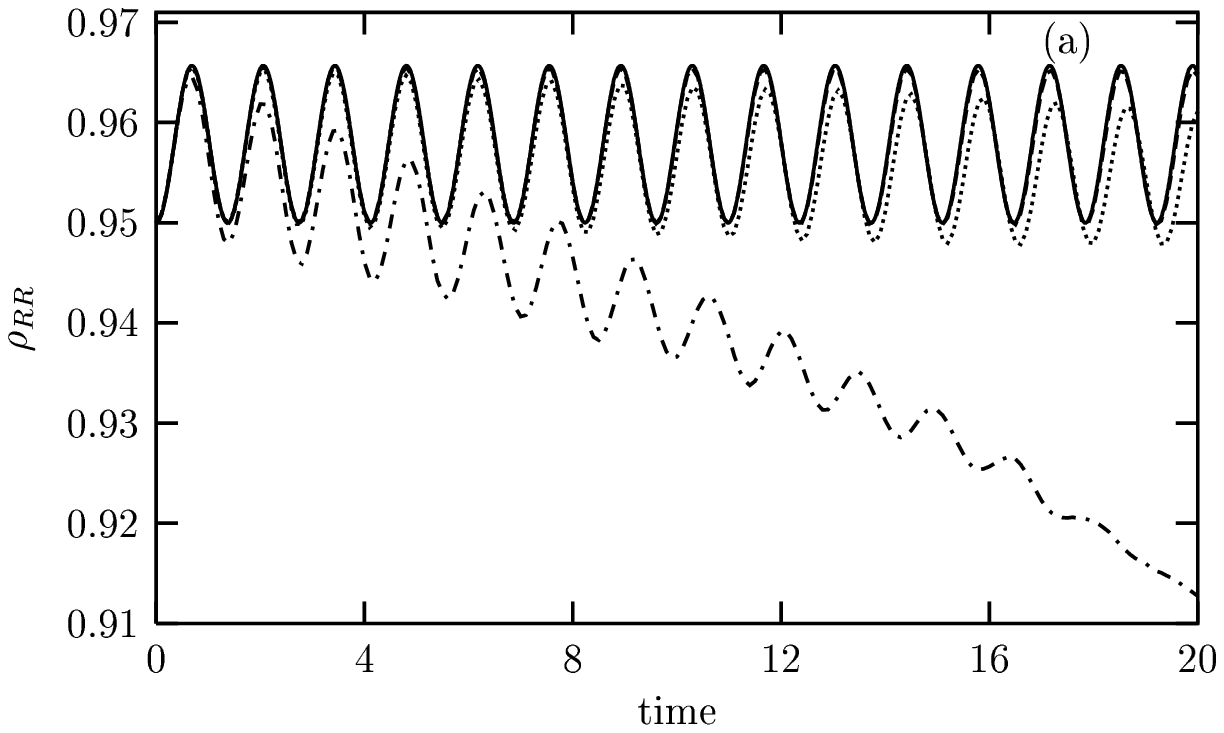,width=0.5\linewidth,clip=} &
\epsfig{file=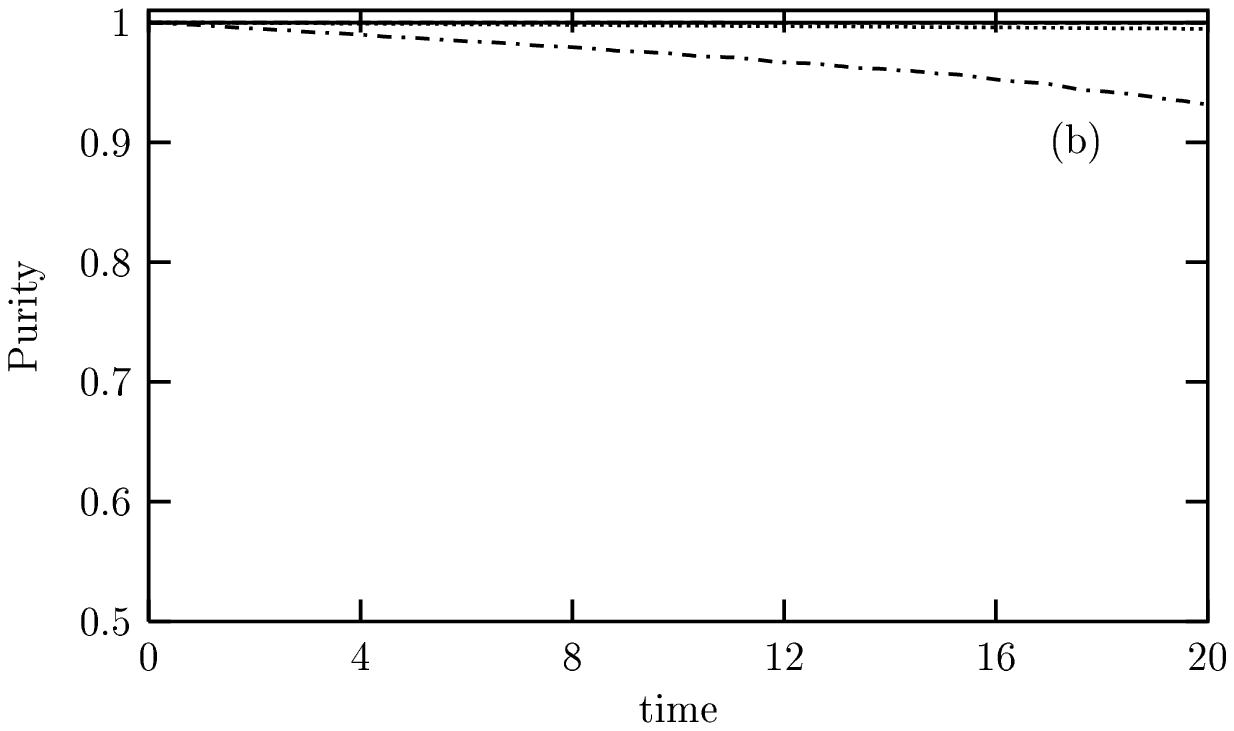,width=0.5\linewidth,clip=} \\
\epsfig{file=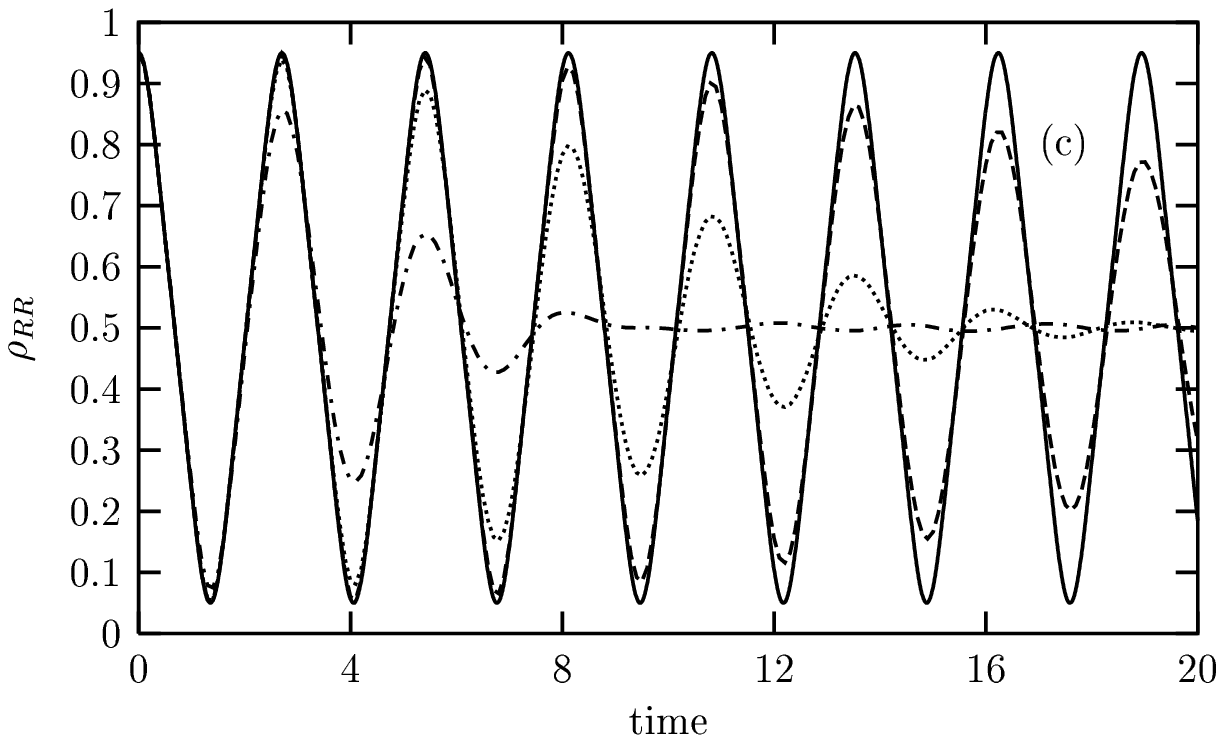,width=0.5\linewidth,clip=} &
\epsfig{file=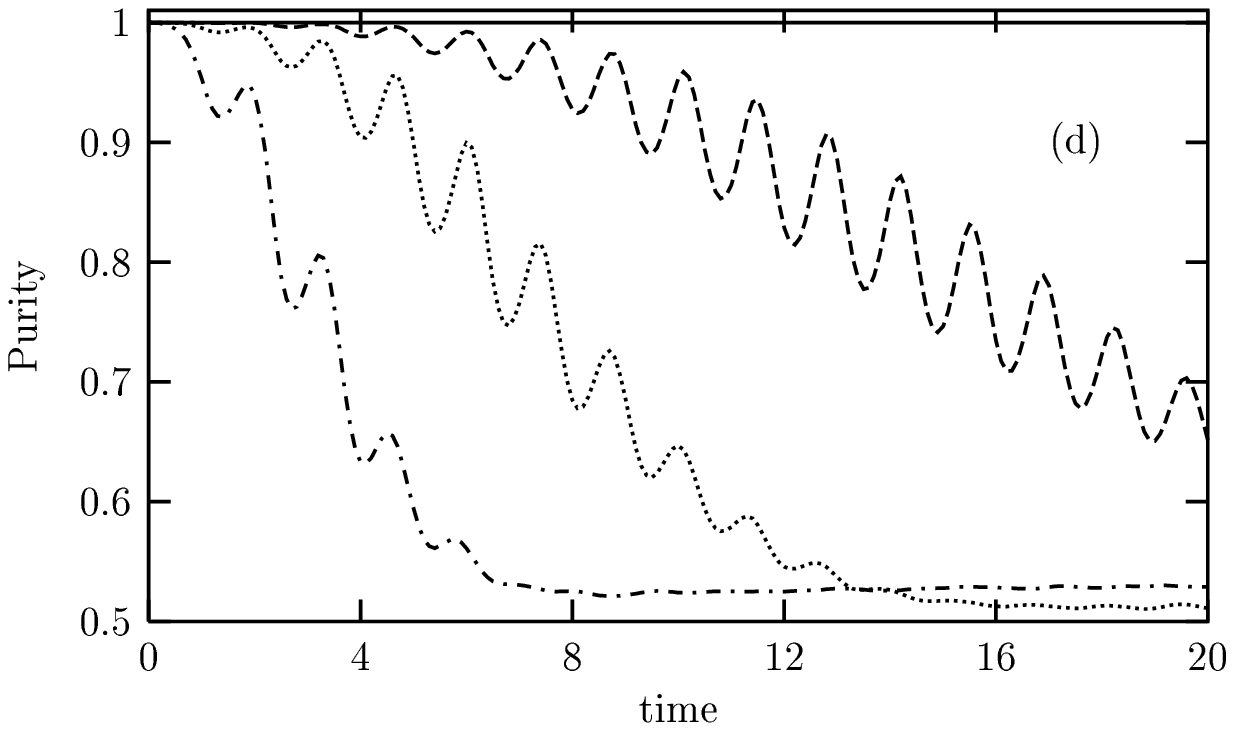,width=0.5\linewidth,clip=}
\end{tabular}
\caption{}
 \label{figv50595}
\end{figure}

\vspace{0.2in} \noindent Fig.  \ref{figv50595}.  $\rho_{RR}$ vs.
the rescaled time in (a) and (c).  Purity vs. the rescaled time in
(b) and (d).  The WDS case with $v=5$ is chosen for (a) and (b) and
the WDS case with $v=-5$ is chosen for (c) and (d). $\Gamma=0$
(thick line), $1.6\times 10^{-4}$ ({\bf - - - -}), $1.6\times
10^{-3}$ ($\cdot\cdot\cdot$), and $1.6\times 10^{-2}$ ($-\cdot-$).
In panels (a) and (b) the $\Gamma = 0$ and $\Gamma = 1.6 \times
10^{-4}$ curves overlap to the extent that they cannot be
distinguished.

\begin{figure}
\centering
\begin{tabular}{cc}
\epsfig{file=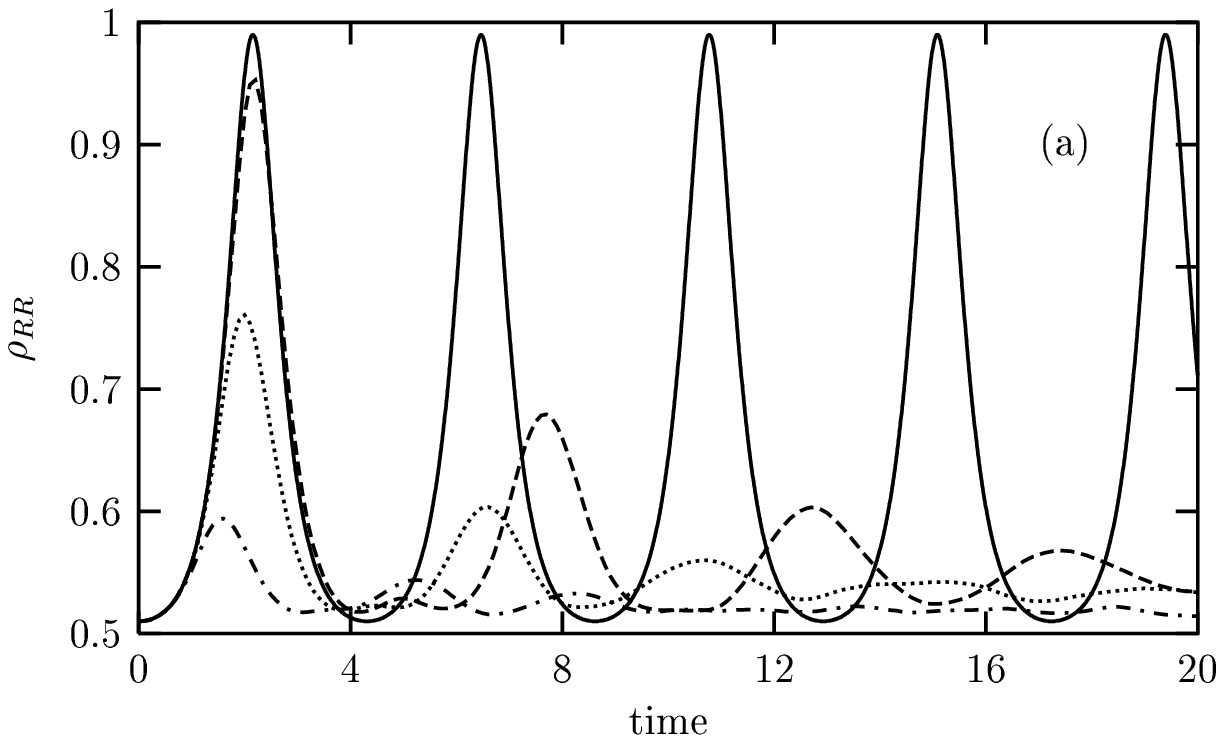,width=0.5\linewidth,clip=} &
\epsfig{file=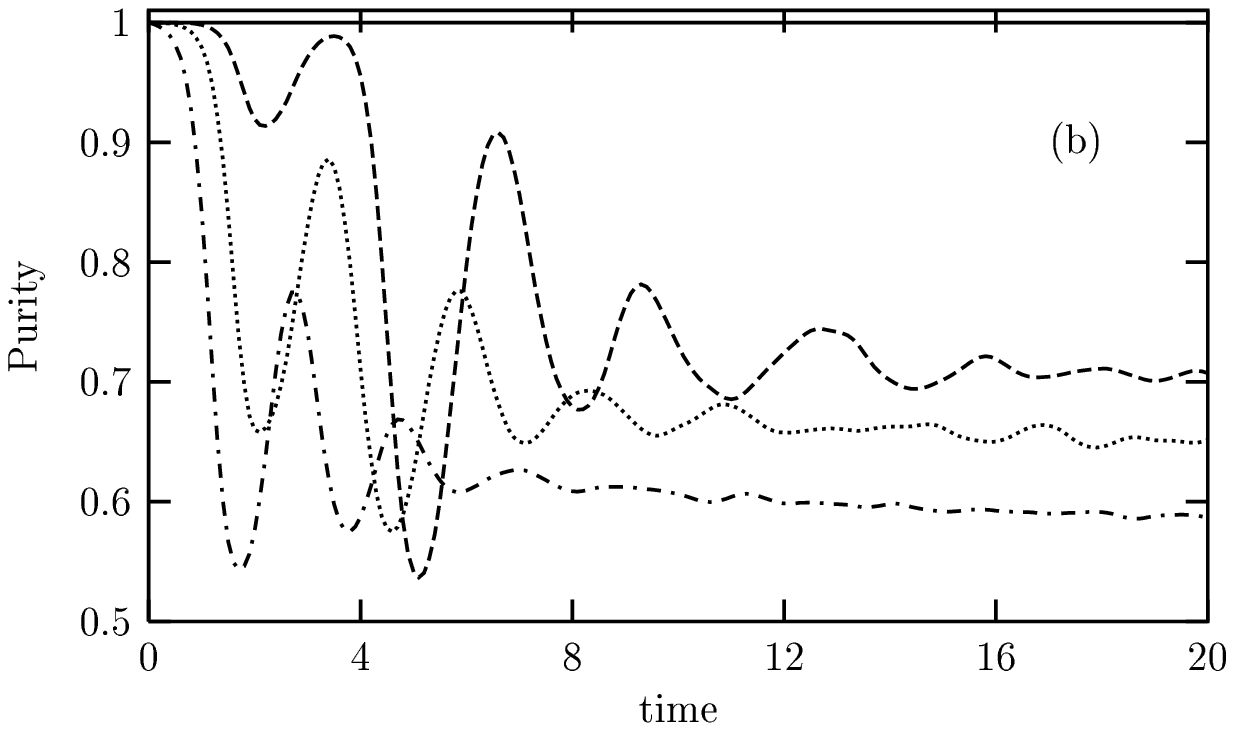,width=0.5\linewidth,clip=} \\
\epsfig{file=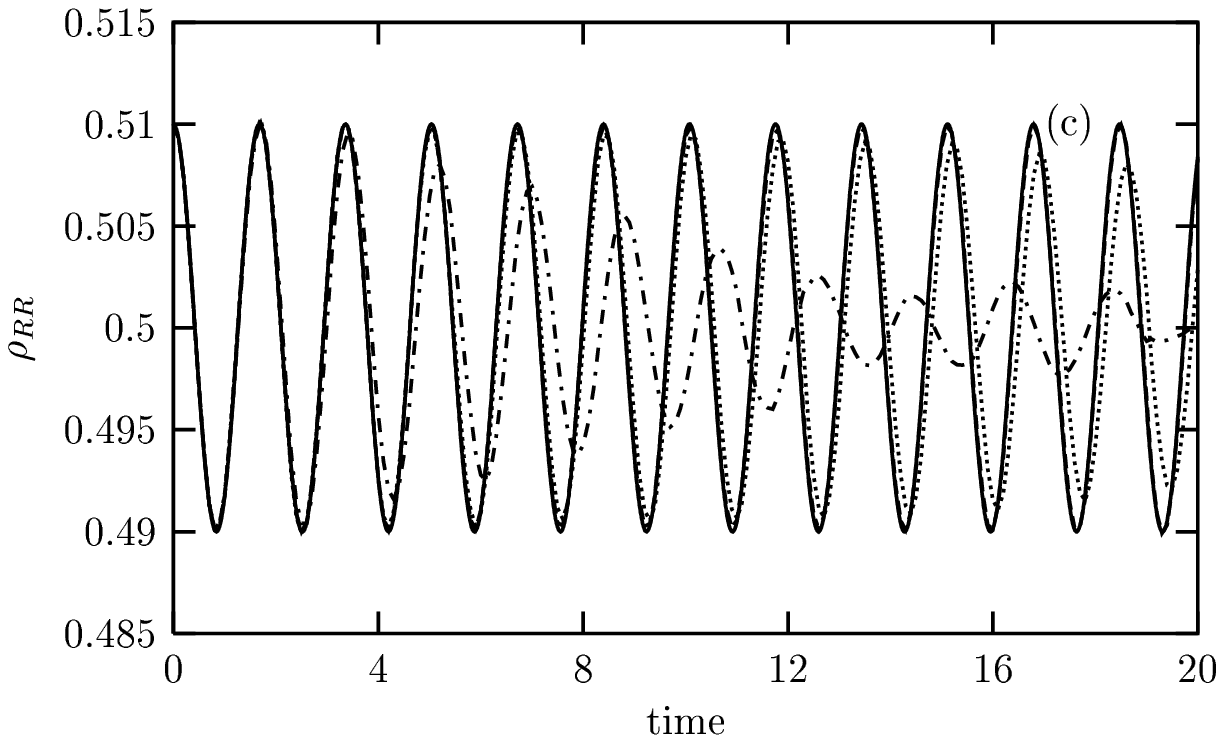,width=0.5\linewidth,clip=} &
\epsfig{file=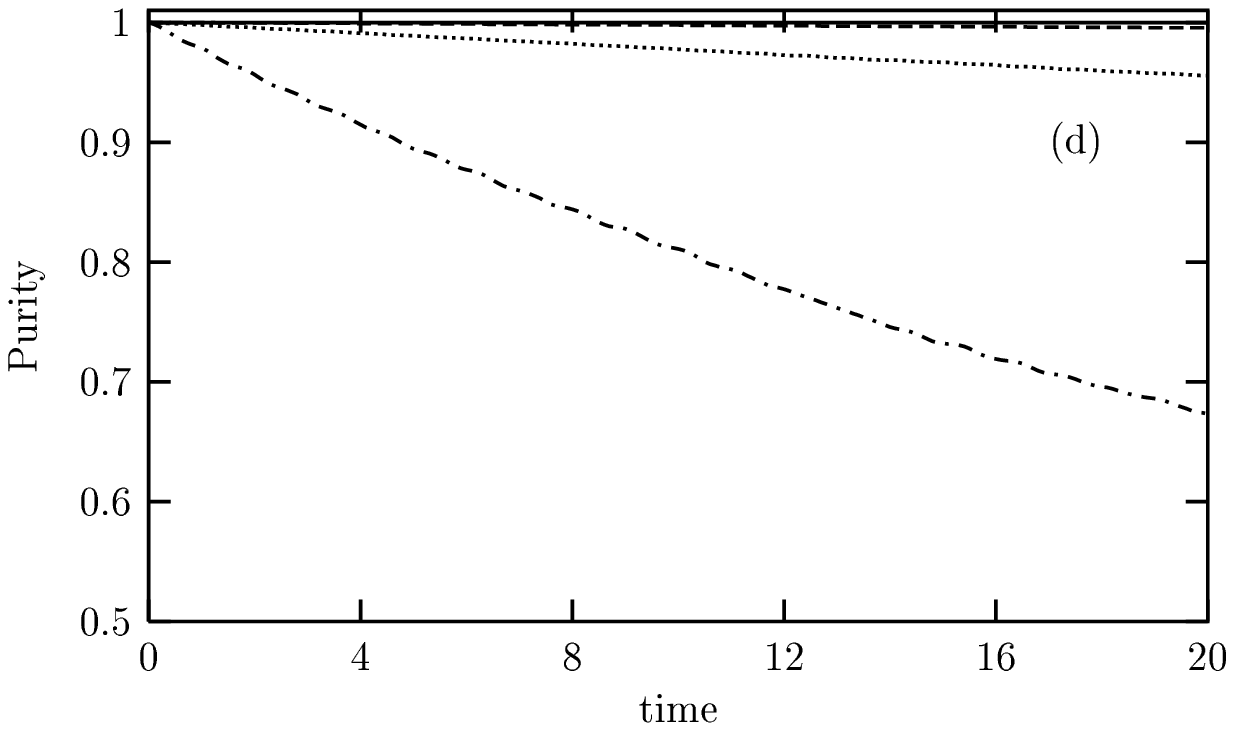,width=0.5\linewidth,clip=}
\end{tabular}
\caption{}
 \label{figv54951}
\end{figure}

\vspace{0.2in} \noindent Fig. \ref{figv54951}. $\rho_{RR}$ vs.
the rescaled time in (a) and (c).  Purity vs. the rescaled time in
(b) and (d). The SDS case with $v=5$ is chosen for (a) and (b) and
the SDS case with $v=-5$ is chosen for (c) and (d). $\Gamma=0$
(thick line), $1.6\times 10^{-4}$ ({\bf - - - -}), $1.6\times
10^{-3}$ ($\cdot\cdot\cdot$), and  $1.6\times 10^{-2} $ ($-\cdot-$).
Note the difference in the scale of ordinate in panels (a) and (c).

\clearpage

\newpage
\begin{thebibliography}{10}
\bibitem{hund}F. Hund, Z. Phys. \textbf{43}, 805 (1927).
\bibitem{sakurai}J. J. Sakurai, \textit{Modern Quantum Mechanics} (Addison-Wesley Publishing
Company, Inc., 1985), Chap 4.
\bibitem{weak_neutral} S. F . Mason, \textit{Molecular Optical Activity and the
Chiral Discriminations}  (Cambridge University Press, Cambridge, 1982),
Chap 11;  G. E. Tranter, Nature (London) \textbf{318}, 172 (1985).
\bibitem{nonlinear1}E. B. Davies, Commun. Math. Phys. \textbf{64}, 191 (1979); A. Koschany, J. Kuffer, G. M. Obermair, and K. Plessner, Phys. Lett.
A \textbf{185}, 412 (1994).
\bibitem{vardi}A. Vardi, J. Chem. Phys. \textbf{112}, 8743 (2000) .
\bibitem{collision}R. Silbey and R. A. Harris, J. Phys. Chem. \textbf{93}, 7062 (1989).
\bibitem{photon}P. Pfeifer, Phys. Rev. A. \textbf{26},701 (1982).
\bibitem{phonon}B. Fain, Phys. Lett. A \textbf{89}, 455 (1982).
\bibitem{harris_cina_romero} J. A. Cina and R. A. Harris, J. Chem. Phys. \textbf{100},
2531 (1994); R. P. Duarte-Zamorano and V. V. Romero-Rochin, J. Chem. Phys. \textbf{114},
9276 (2001).
\bibitem{control_brumer_control_ohtsuki} M. Shapiro, E. Frishman, and P. Brumer, Phys. Rev. Lett.
\textbf{84}, 1669 (2000); Y. Fujimura, L. Gonalez, K. Hoki, J. Manz, and Y. Ohtsuki,
Chem. Phys. Lett. \textbf{306}, 1 (1999).
\bibitem{cplhhan_wardlaw} H. Han and D. M. Wardlaw, Chem. Phys. Lett.
(2008), submitted.
\bibitem{qzeno0}I. O. Stamatescu, E. Joos, H. D. Zeh, C. Kiefer, D. Giulini, and J.
Kupsch, \textit{Decoherence and the Appearance of a Classical World
in Quantum Theory}, 2nd ed. (Springer Verlag, 2003).
\bibitem{hhan_jcp} H. Han and P. Brumer, J. Chem. Phys. \textbf{122},
144316 (2005) and references are therein.
\bibitem{bath_model} E. Joos and H.D. Zeh, Z. Phys. B \textbf{59}, 223 (1985).
\bibitem{bath_model2}A.O. Caldeira and A.J. Leggett, Physica A \textbf{121},
587 (1983).
\bibitem{decoherence_review}  W. H. Zurek, Rev. Mod. Phys. \textbf{75}, 715 (2003); P.
Blanchard, D. Giullini, E. Joos, C. Kiefer, I. O. Stamatescu
(eds), \textit{Decoherence: Theoretical, Experimental, and
Conceptual Problems}, 2nd ed. (Lecture Notes in Physics 538,
Springer Verlag, 2000).
\bibitem{gallis}M. R. Gallis and G. N. Fleming, \pra {\textbf{42}}, 38 (1990).
\bibitem{deco_model} M. J. Gagen, H. M. Wiseman, and G. J. Milburn, \pra {\textbf{48}},
132 (1993); G. J. Milburn, J. Opt. Soc. Am. B \textbf{5}, 1317
(1988).
\bibitem{perci1}N. Gisin and I. Percival, J. Phys. A \textbf{25}, 5677
(1992) and references are therein.
\bibitem{BEC}A. Smerzi, S. Fantoni, S. Giovanazzi, and S. R.
Shenoy, \prl {\textbf{79}}, 4950 (1997).
\bibitem{harris_distinguish}
R. A. Harris, Y. Shi, and J. A. Cina, J. Chem. Phys. {\bf 101},
3459 (1994).
\bibitem{purity} P.C. Lichtner and J.J. Griffin, \prl{\textbf{37}}, 1521 (1976); W.H.
Zurek, S. Habib and J.P. Paz, \prl{\textbf{70}}, 1187 (1993); M.
C. Nemes and A.F.R. deToledo Piza, Physica (Amsterdam)
\textbf{137A}, 367 (1986); X-P. Jiang and P. Brumer, Chem. Phys.
Lett. \textbf{208}, 179 (1993).
\bibitem{torrey} H. C. Torrey, Phys. Rev. \textbf{76}, 1059 (1949).
\bibitem{qoptics}D. F. Walls and G. J. Milburn, \textit{Quantum Optics} (Springer Verlag, 1994); M. O. Scully and M. S. Zubairy, \textit{Quantum optics} (Cambridge University Press, 1997).
\bibitem{sanz_brumer} A. S. Sanz, H. Han, and P. Brumer, J. Chem. Phys. {\bf 124},
214106 (2006).
\bibitem{alkorta} I. Alkorta, O. Picazo, and J. Elguero,
Curr. Org. Chem. \textbf{10}, 695 (2006); J. Ka and S. Shin, J. Mol. Struct. \textbf{623}, 23 (2003).
\end{thebibliography}
\end{document}